%% file: fd_efd.tex
\documentclass[onefignum,onetabnum]{siamart251216}

\usepackage{amssymb}
\usepackage{amsfonts}
\usepackage{graphicx}
\usepackage{epstopdf}
\usepackage{booktabs}
\usepackage{xcolor}
\usepackage{tikz}
\usetikzlibrary{matrix,positioning,decorations.pathreplacing,calc}
\ifpdf
  \DeclareGraphicsExtensions{.pdf,.png,.jpg,.eps}
\else
  \DeclareGraphicsExtensions{.eps}
\fi
\graphicspath{{figures/}}

\newcommand{\vb}[1]{#1}
\newcommand{\mat}[1]{\mathbf{#1}}
\newcommand{\gmat}[1]{\boldsymbol{#1}}
\newcommand{\vx}{\vb{x}}                  %
\newcommand{\xb}{\bar{\vb{x}}}            %
\newcommand{\xp}{\vb{x}'}                 %
\newcommand{\snap}{\vb{\nu}}              %
\newcommand{\stk}{\tilde{\vb{\nu}}}       %
\newcommand{\vz}{\vb{z}}                  %
\newcommand{\vw}{\vb{w}}                  %
\newcommand{\vf}{\vb{f}}                  %
\newcommand{\vg}{\vb{g}}
\newcommand{\vu}{\vb{u}}
\newcommand{\vc}{\vb{c}}
\newcommand{\vphi}{\vb{\phi}}             %
\newcommand{\vpsi}{\vb{\psi}}             %
\newcommand{\vpsit}{\tilde{\vb{\psi}}}    %
\newcommand{\vpsih}{\hat{\vb{\psi}}}      %
\newcommand{\vone}{\mathbf{1}}            %
\newcommand{\mA}{\mat{A}}
\newcommand{\mB}{\mat{B}}
\newcommand{\mC}{\mat{C}}
\newcommand{\mD}{\mat{D}}
\newcommand{\mE}{\mat{E}}
\newcommand{\mF}{\mat{F}}
\newcommand{\mG}{\mat{G}}
\newcommand{\mI}{\mat{I}}
\newcommand{\mJ}{\mat{J}}
\newcommand{\mL}{\mat{L}}
\newcommand{\mM}{\mat{M}}
\newcommand{\mO}{\mat{O}}
\newcommand{\mP}{\mat{P}}
\newcommand{\mQ}{\mat{Q}}
\newcommand{\mR}{\mat{R}}
\newcommand{\mS}{\widetilde{\mat{S}}}       %
\newcommand{\mU}{\mat{U}}
\newcommand{\mW}{\mat{W}}
\newcommand{\mV}{\mat{V}}                %
\newcommand{\mVt}{\widetilde{\mat{V}}}      %
\newcommand{\Vx}{\mV_{1:K-1}}           %
\newcommand{\Vy}{\mV_{2:K}}             %
\newcommand{\Vxt}{\mVt_{1:K-1}}
\newcommand{\Vyt}{\mVt_{2:K}}
\newcommand{\mPhi}{\gmat{\Phi}}           %
\newcommand{\mSigma}{\gmat{\Sigma}}
\newcommand{\mLambda}{\gmat{\Lambda}}
\newcommand{\mTheta}{\gmat{\Theta}}
\newcommand{\mPi}{\gmat{\Pi}}
\newcommand{\mObs}{\boldsymbol{\mathcal{O}}}  %
\newcommand{\CC}{\mathbb{C}}
\newcommand{\RR}{\mathbb{R}}
\newcommand{\ZZ}{\mathbb{Z}}

\newcommand{\ii}{\mathrm{i}}
\newcommand{\ee}{\mathrm{e}}
\newcommand{\dd}{\mathrm{d}}
\newcommand{\dt}{\Delta t}
\newcommand{\herm}{\mathsf{H}}
\newcommand{\trans}{\top}
\newcommand{\pinv}{\dagger}
\newcommand{\abs}[1]{\lvert #1\rvert}
\newcommand{\norm}[1]{\lVert #1\rVert}
\newcommand{\Kry}{\mathcal{K}}            %
\newcommand{\Jset}{\mathcal{J}}           %
\newcommand{\muh}{\hat{\mu}}              %
\newcommand{\betah}{\hat{\beta}}          %
\newcommand{\nt}{\tilde{n}}              %
\newcommand{\rhoh}{\hat{\rho}}            %
\DeclareMathOperator{\spec}{spec}
\DeclareMathOperator{\rank}{rank}
\DeclareMathOperator{\range}{range}
\DeclareMathOperator{\spn}{span}
\DeclareMathOperator{\diag}{diag}
\DeclareMathOperator{\blkdiag}{blkdiag}
\DeclareMathOperator{\re}{Re}
\DeclareMathOperator{\im}{Im}

\headers{Floquet and ensemble Floquet decomposition}{J. M. Quinlan}

\title{Floquet Decomposition: Floquet Multipliers and Time-Periodic Modes from Period-Sampled Snapshot Data\thanks{Submitted to the editors November 25, 2026.
\funding{This work was supported by the Air Force Office of Scientific Research under award FA9550-14-1-0051 (program officer Dr.\ Mitat Birkan).}}}

\author{J. Matt Quinlan\thanks{Department of Mechanical and Aerospace Engineering, University of Colorado Colorado Springs, Colorado Springs, CO 80918 USA (\email{jquinlan@uccs.edu}).}}

\begin{document}

\maketitle

\begin{abstract}
Dynamic mode decomposition (DMD) fits one constant linear operator to a sequence of snapshots.
When the dynamics, or the view of them, vary periodically, as in a parametrically excited flow or a rotating structure seen in a fixed plane, no such operator exists, and stability is decided by the Floquet multipliers, the eigenvalues of the map over one period.
In the paper that introduced DMD, Schmid observed that snapshots sampled at the period of an oscillatory flow would identify the Floquet matrix.
We develop that observation into a method and analyze it.
The Floquet decomposition (FD) applies DMD to fixed-rate data subsampled once per period; the ensemble Floquet decomposition (EFD) stacks the snapshots of each period into one column, giving one set of multipliers and phase-resolved modes.
For linear time-periodic systems both are exact: FD when the measurement at one instant determines the state, EFD when the measurements over one period do---so EFD succeeds with partial measurements.
EFD and Hankel (delay-embedded) DMD use the same data matrix with different shifts between columns.
With $p$ samples per period, Hankel-DMD returns each Floquet mode as a carrier with sidebands, a comb of $p$-th roots of its multiplier, which the period shift collapses onto the multiplier.
The methods are verified against known multipliers; outside their hypotheses, on nonlinear or noisy data, the estimated moduli are biased and are not decay rates.
Measurements of velocity, scalars, and acoustic pressure in a swirl-stabilized combustor illustrate the choice of period, rank, and data scaling.
\end{abstract}



\input{sections/01_introduction}
\input{sections/02_background}
\input{sections/03_method}
\input{sections/04_relations}
\input{sections/05_verification}
\input{sections/06_symmetry}
\input{sections/07_experiment}
\input{sections/08_guidance}
\input{sections/09_conclusions}

\appendix
\input{sections/A_proofs}

\section*{Acknowledgments}
The author thanks Professor Ben T.\ Zinn, who advised the dissertation in which this work began, and Professor Luca Dieci, who pointed the way to the theory of linear periodic systems.
The author also thanks Dr.\ Mitat Birkan of the Air Force Office of Scientific Research for supporting the work under award FA9550-14-1-0051.
This work used the INCLINE cluster at the University of Colorado Colorado Springs; INCLINE is supported by the National Science Foundation, grant \#2017917.
A large language model (Claude, Anthropic) was used to edit the text and to draft the proofs of \cref{app:proofs}.
The author assumes responsibility for all content.
The development reported here was carried out in 2013--2014 and first documented in \cite{Quinlan-2015}.

\bibliographystyle{siamplain}
\bibliography{references}

\end{document}

%% file: sections/01_introduction.tex
\section{Introduction}
\label{sec:introduction}

Time-resolved measurements and simulations now routinely resolve the dynamics of complex flows, yet turning these records into statements about the stability of the observed motion, rather than descriptions of it, remains difficult when the governing equations are unknown or intractable.
Combustion instability is a case in point: the quantity on which the design of a combustor depends is a stability margin, and what the experiment provides is a sequence of images and acoustic pressure traces \cite{Lieuwen-2005,Culick-2006}.

Dynamic mode decomposition (DMD) \cite{Schmid-2010} has become a standard tool for such data because it requires nothing but observations (snapshots), returns modes that each have a single frequency and growth rate, and inherits an interpretation from the spectral theory of the Koopman operator \cite{Rowley-2009,Mezic-2005,Mezic-2013}.
Its many extensions \cite{Tu-2014,Jovanovic-2014,Proctor-2016,Williams-2015,Askham-2018,Kutz-2016-mrDMD}, among them the delay-embedded forms known as higher-order and Hankel-DMD \cite{LeClainche-2017,Arbabi-2017,Brunton-2017}, are surveyed by Kutz et al.~\cite{Kutz-2016-book} and Schmid~\cite{Schmid-2022}.

Every one of these methods fits a single, constant linear map to the chosen observables, $\snap_{k+1}\approx\mA\snap_k$, where $\snap_k$ is the $k$-th snapshot, and the assumption can fail for four distinct reasons.
Nonlinearity is a problem of closure: no finite-dimensional linear model whose observables include the state can represent a system with more than one isolated invariant set \cite{Brunton-2016-PLoS}.
Noise and finite records bias the eigenvalues \cite{Arbabi-2017,Dawson-2016,Hemati-2017}.
Partial observation, as when a three-dimensional flow is measured in a plane, leaves the sequence $\snap_k$ without a constant map even for linear dynamics, and the remedy is delay embedding \cite{Takens-1981,Arbabi-2017,Kamb-2020}.
The fourth, the subject of this paper, is a time-dependent operator: a linear system $\dot{\vx} = \mA(t)\vx$, observed in full, has no constant evolution operator, and the eigenvalues of $\mA(t)$ at any instant say nothing reliable about its stability \cite{Chicone-2011,Markus-1960}.
Windowed and online forms of DMD \cite{Zhang-2019} track operators that vary slowly, and phase averaging \cite{Reynolds-1972,Heidt-2024} characterizes flows with periodic statistics, but neither returns the stability of a system whose operator varies periodically on the time scale of the dynamics.

Many flows that defeat a constant-operator description are, however, periodic or nearly so, and for linear time-periodic operators a complete stability theory has existed since Floquet~\cite{Floquet-1883}: over one period the state is advanced by a constant matrix, the monodromy matrix, whose eigenvalues, the Floquet multipliers, decide stability, a perturbation growing if any multiplier lies outside the unit circle \cite{Hale-1980}.
When the governing equations are available, Floquet analysis is a standard tool of fluid mechanics, used for flows that are driven periodically \cite{Davis-1976}, for the Faraday instability of a vibrated interface \cite{Benjamin-1954,Kumar-1994}, and for the secondary instabilities of boundary layers and bluff-body wakes, whose base states are themselves periodic \cite{Herbert-1988,Barkley-1996}.
The canonical example is the Mathieu equation, in which the periodic variation of a parameter, rather than any forcing of the state, destabilizes an oscillator whose instantaneous eigenvalues are all neutral \cite{Kovacic-2018}; flames in an acoustic field are a combustion example \cite{Searby-1991,Assier-2014}.
The question addressed here is how to obtain the multipliers from measurements, when there is no operator to linearize.

The periodicity may reside in either of two places.
If the data are generated by $\vx_{k+1} = \mA_k\vx_k$ and observed through $\snap_k = \mC_k\vx_k$, a constant-operator fit to $\snap_k$ fails when either the dynamics $\mA_k$ or the measurement $\mC_k$ is periodic, and the two cases are equivalent up to a periodic change of coordinates, so the data cannot distinguish them.
Physically, periodic dynamics arise from parametric excitation, as when a coherent oscillation of one subsystem modulates the properties of another.
Periodic observation arises when an autonomous flow with a rotational symmetry has a rotating wave as its base state, such as the precessing vortex core of a swirling jet \cite{Syred-2006}: the full state could be viewed from a co-rotating frame in which nothing is periodic, but a measurement in a fixed plane cannot.

The possibility of extracting Floquet multipliers from snapshot data was anticipated in the paper that introduced DMD.
In a discussion of sampling rates, Schmid~\cite[\S\,2.6]{Schmid-2010} observed that
\begin{quote}
``If the sampling frequency is tuned to the frequency of an oscillatory flow, the mapping from period to period will identify the Floquet matrix whose eigenvalues represent the Floquet multipliers.''
\end{quote}
The remark was not pursued there.
Mezi\'c and Surana~\cite{Mezic-2016} subsequently established the Koopman-theoretic structure of time-periodic systems, in which the Floquet exponents $\rho_l$ generate a lattice of Koopman eigenvalues $\rho_l + \ii j\omega$, with $\omega$ the fundamental frequency and $j$ any integer, and sketched a computational procedure that applies DMD phase by phase to ensembles of trajectories, whose validation they left to future work.
Within the DMD and Koopman literature, an estimator of the multipliers from a single trajectory, with conditions under which it is exact and a comparison against known multipliers, has to our knowledge not been given; the closest counterparts are the subspace identification methods for periodic systems developed in control theory and structural dynamics (\cref{sec:other-methods}).

In this paper we develop Schmid's observation into a practical method, the Floquet decomposition (FD), and its multi-phase extension, the ensemble Floquet decomposition (EFD), which estimate the leading Floquet multipliers of a time-periodic system, and with them its stability, from a single record of snapshot data, without knowledge of the governing operator.
FD keeps data recorded at a fixed rate and subsamples it once per period, so that the period can be chosen after the measurement; the subsampled snapshots are iterates of the monodromy matrix.
EFD stacks all the snapshots of one period into each column of the data matrix, so that every phase of the period shares one set of multipliers and the modes are resolved in phase.

The paper is written for the reader who wishes to use the method and needs to know when it can be trusted.
The argument rests on one fact of linear algebra: DMD returns exactly the eigenvalues of a constant matrix when the vectors it is given are successive iterates of that matrix, seen through a measurement that loses none of them (\cref{sec:dmd}); period-sampled snapshots of a linear periodic system are such iterates of the monodromy matrix.
The contributions built on this are the following.
\begin{enumerate}
\item[(i)] \emph{Conditions for exactness} (\cref{sec:methods}).
FD requires that the measurement at one instant determine the excited state; EFD requires only that the measurements made during one period determine it, and this, rather than any gain in rank, is its advantage for cases with partial measurements.
The EFD mode is the Floquet mode at each phase of the period.
\item[(ii)] \emph{Relation to Hankel-DMD and to the Koopman spectrum} (\cref{sec:relations}).
On the same delay-embedded data, Hankel-DMD returns for each multiplier a comb of its $p$-th roots, one for each harmonic present in the observed Floquet mode, which is the sampled form of the lattice of Mezi\'c and Surana~\cite{Mezic-2016}; the period shift collapses the comb onto the multipliers.
The identities underneath are the lifted and cyclic reformulations of periodic systems known in control theory \cite{Meyer-1975,Bittanti-2009}; what is established here is their consequence for the two data-driven methods.
\item[(iii)] \emph{Verification} (\cref{sec:verification}) against the known multipliers of the Mathieu equation and of a limit cycle, and of what the methods return when their hypotheses fail: for data far from a linear periodic system the moduli are biased and are not decay rates.
\item[(iv)] \emph{Symmetry and observation} (\cref{sec:symmetry}): for a rotating wave observed in a fixed frame, the multipliers are the exponentials of the co-rotating eigenvalues over one period, up to a root of unity.
\item[(v)] \emph{Application and practice} (\cref{sec:experiment,sec:guidance}): on simultaneous velocity, scalar, and dynamic pressure measurements from a swirl-stabilized combustor, DMD and FD miss the two modes associated with the instability and EFD and Hankel-DMD find both.
The combustion physics is treated in a companion paper \cite{Quinlan-2026-PCI}, which suggests that the instability is a parametric, acoustically coupled combustion instability.
\end{enumerate}
Formal statements and proofs are collected in \cref{app:proofs}.

%% file: sections/02_background.tex
\section{Time-periodic systems and their sampled data}
\label{sec:floquet}

This section collects the parts of Floquet theory that the method uses \cite{Floquet-1883,Hale-1980,Chicone-2006} and sets out the model of the sampled data on which the later results rest.
Readers familiar with Floquet theory may pass to \cref{sec:sampled}.

\subsection{Floquet theory in brief}
\label{sec:floquet-theory}

Consider the linear system
\begin{equation}
  \dot{\vx} = \mA(t)\,\vx, \qquad \vx(0) = \vx_0 \in \CC^n, \qquad \mA(t+T) = \mA(t),
  \label{eq:ltp}
\end{equation}
with $T>0$ the period and $\omega = 2\pi/T$.
In the applications we have in mind $\vx$ is a perturbation of a base flow and $\mA(t)$ the operator linearized about it, periodic either because the base flow is periodic or because a parameter is modulated.

When $\mA$ is constant the solution is $\vx(t) = \ee^{\mA(t-t_0)}\vx(t_0)$, and the eigenvalues of $\mA$ decide stability.
When $\mA$ depends on time the solution is still advanced linearly,
\begin{equation}
  \vx(t) = \mPhi(t,t_0)\,\vx(t_0), \qquad \partial_t\mPhi(t,t_0) = \mA(t)\,\mPhi(t,t_0), \quad \mPhi(t_0,t_0) = \mI,
  \label{eq:stm}
\end{equation}
where the state-transition matrix $\mPhi(t,t_0)$ carries the state at the time $t_0$ to the state at the time $t$.
It is in general not the exponential of anything simple, and it depends on $t$ and $t_0$ separately, not only on the elapsed time $t-t_0$.
Two of its properties are used throughout: advancing from $t_0$ to an intermediate time $t_1$ and then from $t_1$ to $t$ is the same as advancing from $t_0$ to $t$, $\mPhi(t,t_1)\mPhi(t_1,t_0) = \mPhi(t,t_0)$; and, because $\mA$ is periodic, shifting both times by one period changes nothing, $\mPhi(t+T,t_0+T) = \mPhi(t,t_0)$.
The second property says that the map over one full period is the same whichever period it is applied to.
That map is the \emph{monodromy matrix},
\begin{equation}
  \mM = \mPhi(T,0),
  \label{eq:monodromy}
\end{equation}
and it follows that
\begin{equation}
  \vx(kT) = \mM^{k}\,\vx_0, \qquad k = 0,1,2,\dots
  \label{eq:strobe}
\end{equation}
Viewed once per period, as under a stroboscope, a time-periodic linear system is a time-invariant one: the same constant matrix is applied at every step.
Everything in this paper is built on \cref{eq:strobe}.

The eigenvalues $\mu_1,\dots,\mu_n$ of $\mM$ are the \emph{Floquet multipliers}; none is zero, since $\mM$ is invertible.
By \cref{eq:strobe} a perturbation along the eigenvector $\vw_l$ of $\mM$ is multiplied by $\mu_l$ every period, so the zero solution of \cref{eq:ltp} is asymptotically stable if and only if $\abs{\mu_l}<1$ for all $l$, and it is unstable if $\abs{\mu_l}>1$ for some $l$.
A number $\rho_l$ with
\begin{equation}
  \mu_l = \ee^{\rho_l T}
  \label{eq:exponent}
\end{equation}
is a \emph{Floquet exponent}; its real part $\ln\abs{\mu_l}/T$ is a growth rate and its imaginary part a frequency.
The multipliers are uniquely determined by \cref{eq:ltp}, but each exponent is determined only up to an integer multiple of $\ii\omega$: if $\rho_l$ satisfies \cref{eq:exponent}, so does $\rho_l + \ii j\omega$ for every integer $j$.
The reason is that of the stroboscope, which cannot tell how many whole cycles have passed between two flashes.

These are statements about $\mM$, not about $\mA(t)$ at any instant.
As Chicone puts it, ``there is no obvious relation between the (time-dependent) eigenvalues of a time-dependent matrix and the stability of the zero solution of the corresponding first-order linear differential equation'' \cite[pp.~1658--1659]{Chicone-2011}, and the example of Markus and Yamabe~\cite{Markus-1960} is unstable although the eigenvalues of $\mA(t)$ have negative real parts for every $t$.

Between the stroboscopic instants the solution has a simple structure.
Floquet's theorem states that $\mPhi(t,0) = \mP(t)\,\ee^{\mB t}$ with $\mP(t+T)=\mP(t)$, $\mP(0)=\mI$, and $\mB$ a constant, in general complex, matrix, so that $\mM = \ee^{\mB T}$ and the change of variables $\vx = \mP(t)\vz$ reduces \cref{eq:ltp} to the constant-coefficient system $\dot{\vz} = \mB\vz$.
For an eigenpair $\mM\vw_l = \mu_l\vw_l$ and a choice of exponent $\rho_l$, define the \emph{Floquet mode}
\begin{equation}
  \vphi_l(t) = \ee^{-\rho_l t}\,\mPhi(t,0)\,\vw_l .
  \label{eq:floquet-mode}
\end{equation}
Since $\mPhi(t+T,0) = \mPhi(t+T,T)\mPhi(T,0) = \mPhi(t,0)\mM$, the mode is $T$-periodic.
If $\mM$ is diagonalizable and $\vx_0 = \sum_l a_l\vw_l$, then
\begin{equation}
  \vx(t) = \sum_{l=1}^{n} a_l\, \ee^{\rho_l t}\, \vphi_l(t) .
  \label{eq:floquet-expansion}
\end{equation}
This is to be compared with the expansion $\vx(t) = \sum_l a_l\ee^{\lambda_l t}\vw_l$ of a system with constant $\mA$: the shape of each mode, instead of being fixed, changes periodically while the mode grows or decays.

A periodic shape has a Fourier series, $\vphi_l(t) = \sum_j\hat{\vphi}_{l,j}\,\ee^{\ii j\omega t}$, so in a spectrum a Floquet mode is not one line but a carrier with sidebands at the frequencies $\im\rho_l + j\omega$, all of which grow or decay at the one rate $\re\rho_l$; this is the source of the difficulty that time-periodic systems present to DMD, which assigns one eigenvalue to each line, and of the structure found in \cref{sec:relations}.
Replacing $\rho_l$ by $\rho_l + \ii j\omega$ replaces $\vphi_l(t)$ by $\ee^{-\ii j\omega t}\vphi_l(t)$, which is again $T$-periodic: the branch of the logarithm in \cref{eq:exponent} decides which spectral line is called the carrier, and only the product $\ee^{\rho_l t}\vphi_l(t)$ is intrinsic (\cref{sec:efd-modes,sec:lattice}).

\begin{figure}
\centering
\includegraphics[width=0.86\textwidth]{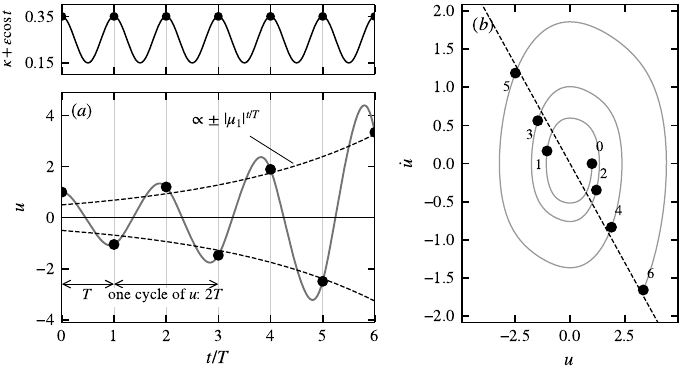}
\caption{The stroboscopic view of a time-periodic system: the Mathieu equation \cref{eq:mathieu} with $\kappa = 0.25$, $\varepsilon = 0.1$, started at $(u,\dot u) = (1,0)$.
(a) The modulated stiffness (top) and the solution (bottom, gray), with their samples once per period $T$ of the modulation (circles).
The response oscillates with the period $2T$, so the samples alternate in sign; the dashed envelope grows as $\abs{\mu_1}^{t/T}$.
(b) The same solution in the phase plane; the samples, numbered by period, are the iterates $\mM^k\vx_0$ and approach the eigenvector $\vw_1$ of the monodromy matrix (dashed), on alternate sides of the origin.}
\label{fig:strobe}
\end{figure}

The canonical example is the Mathieu equation
\begin{equation}
  \ddot{u} + \left(\kappa + \varepsilon\cos t\right)u = 0 ,
  \label{eq:mathieu}
\end{equation}
which has the form \cref{eq:ltp} with $\vx = (u,\dot u)^{\trans}$ and $T = 2\pi$ and which models an oscillator whose stiffness, rather than its forcing, is modulated \cite{Magnus-2013,Kovacic-2018,Verhulst-2011}.
For small $\varepsilon$ it is unstable in narrow tongues that emanate from $\kappa = j^2/4$, $j=1,2,\dots$, even though the eigenvalues of $\mA(t)$ are purely imaginary at every instant.
\Cref{fig:strobe} shows a solution in the first tongue, where the multipliers are real and negative, $\mu_1 = -1.367$ and $\mu_2 = -0.731$.
The period $T$ is that of the modulation, not of the response, which oscillates at half its frequency, so the samples taken once per period alternate in sign, which is what a negative multiplier means, grow by the factor $\abs{\mu_1}$, and settle onto the eigenvector $\vw_1$ of the monodromy matrix.
Sampling every $2T$ instead would give the positive multipliers $\mu_l^2$ and the same verdict on stability, but would hide the subharmonic response.
The samples obey \cref{eq:strobe}; the continuous trajectory between them obeys no constant-coefficient equation.

\subsection{Sampled data: periodic dynamics or periodic observation}
\label{sec:sampled}

Data are recorded at a fixed interval $\dt$.
Throughout the analysis we assume \emph{commensurate sampling}, a period that is a whole number of sampling intervals,
\begin{equation}
  T = p\,\dt, \qquad p \in \ZZ,\ p \ge 1,
  \label{eq:commensurate}
\end{equation}
so that $p$ is the number of samples per period; \cref{sec:period} treats the case in which $T/\dt$ is not an integer or $T$ is known only approximately.
With $t_k = k\dt$, $\vx_k = \vx(t_k)$, and $\mPhi_k = \mPhi(t_{k+1},t_k)$, the samples obey
\begin{equation}
  \vx_{k+1} = \mPhi_k\, \vx_k , \qquad \snap_k = \mC_k\, \vx_k \in \CC^m , \qquad \mPhi_{k+p} = \mPhi_k , \quad \mC_{k+p} = \mC_k .
  \label{eq:sampled}
\end{equation}
Here $\snap_k$ is the snapshot, the vector of the $m$ quantities measured at time $t_k$ (the pixels of an image, the vectors of a velocity field, the signals of a set of sensors), and the measurement matrix $\mC_k\in\CC^{m\times n}$ says how they depend on the state.
A record consists of $N$ snapshots, $\snap_1,\dots,\snap_N$, the first taken one sampling interval after the initial time.
In an experiment $\mC_k$ is seldom known and seldom invertible: a planar measurement of a three-dimensional flow sees the state only in part.
The methods below use neither $\mC_k$ nor $\mPhi_k$; the model \cref{eq:sampled} serves only to say what the data are assumed to be.

We allow the measurement matrix to be periodic as well as the dynamics, because either source of periodicity defeats a constant-operator description of the data, and because the two cannot be told apart from the data.
By Floquet's theorem the substitution $\vx = \mP(t)\vz$ turns \cref{eq:sampled} into $\vz_{k+1} = \ee^{\mB\dt}\vz_k$, $\snap_k = \bigl(\mC_k\mP(t_k)\bigr)\vz_k$: a time-invariant system observed through a periodic measurement.
Conversely, a time-invariant system observed through a periodic $\mC_k$ is of the form \cref{eq:sampled}.
The sequence of snapshots is identical in the two descriptions, and so is everything computed from it.
Whether the periodicity in a data set is parametric (in the dynamics) or observational is therefore not decidable from the data; it requires knowledge of the physics.
The Floquet multipliers are unaffected by the ambiguity, because they are invariant under periodic changes of coordinates.

The stroboscope may be set to flash at any phase of the period, and the monodromy matrix depends on that phase.
Define the monodromy matrix at phase $i$ by
\begin{equation}
  \mM_i = \mPhi(t_i+T,\,t_i) = \mPhi_{i+p-1}\cdots\mPhi_{i+1}\mPhi_i , \qquad i = 1,\dots,p,
  \label{eq:Mi}
\end{equation}
so that the last phase is the initial one, $\mM_p = \mPhi(2T,T) = \mM$, and
\begin{equation}
  \vx_{i+kp} = \mM_i^{\,k}\, \vx_i , \qquad k = 1,2,\dots
  \label{eq:period-map}
\end{equation}
Since $\mM_i\,\mPhi(t_i,0) = \mPhi(t_i+T,0) = \mPhi(t_i+T,T)\,\mPhi(T,0) = \mPhi(t_i,0)\,\mM$, the matrices at different phases are similar,
\begin{equation}
  \mM_i = \mPhi(t_i,0)\, \mM\, \mPhi(t_i,0)^{-1} .
  \label{eq:Mi-similar}
\end{equation}
They therefore have the same eigenvalues, with the same multiplicities, but not the same eigenvectors: if $\mM\vw_l = \mu_l\vw_l$ then
\begin{equation}
  \mM_i\, \vw_l^{(i)} = \mu_l\, \vw_l^{(i)} , \qquad \vw_l^{(i)} = \mPhi(t_i,0)\,\vw_l = \ee^{\rho_l t_i}\,\vphi_l(t_i) .
  \label{eq:Mi-eigvec}
\end{equation}
The multipliers are independent of the phase; the eigenvector of $\mM_i$ is the Floquet mode evaluated at phase $t_i$, scaled by $\ee^{\rho_l t_i}$.
This is the reason period-sampled data from different phases can be combined into a single eigenvalue problem (\cref{sec:efd}) and the reason their eigenvector blocks trace out a time-periodic mode.

Two further objects appear throughout.
The first is the \emph{stacked snapshot}, or delay vector, of window length $d$, which collects $d$ consecutive snapshots into one tall vector.
The second is the matrix that relates it to the state at the start of the window, the \emph{observability matrix over the window}:
\begin{equation}
  \stk_k^{(d)} = \begin{bmatrix} \snap_k \\ \snap_{k+1} \\ \vdots \\ \snap_{k+d-1} \end{bmatrix} \in \CC^{md},
  \qquad
  \mObs_k^{(d)} = \begin{bmatrix} \mC_k \\ \mC_{k+1}\,\mPhi(t_{k+1},t_k) \\ \vdots \\ \mC_{k+d-1}\,\mPhi(t_{k+d-1},t_k) \end{bmatrix} \in \CC^{md\times n},
  \label{eq:delay-vector}
\end{equation}
\begin{equation}
  \stk_k^{(d)} = \mObs_k^{(d)}\, \vx_k , \qquad \mObs_{k+p}^{(d)} = \mObs_k^{(d)} .
  \label{eq:h-equals-Ox}
\end{equation}
Each block of $\mObs_k^{(d)}$ carries the state forward to a later instant and measures it there.
We say that \cref{eq:sampled} is \emph{observable over one period} from phase $i$ on a subspace $\mathcal{S}\subseteq\CC^n$ if $\mObs_i^{(p)}$ is one-to-one (injective) on $\mathcal{S}$: no non-zero state in $\mathcal{S}$ produces zero output throughout a period, or, equivalently, a state in $\mathcal{S}$ can be reconstructed from the measurements made during one period.
For $d=1$ the condition is that the single matrix $\mC_i$ be one-to-one on $\mathcal{S}$, which is much stronger: the state must be recoverable from the measurement at one instant.

\subsection{Periodic orbits of nonlinear systems}
\label{sec:nonlinear-background}

Let $\xb(t)$ be a $T$-periodic solution of $\dot{\vx} = \vf(\vx,t)$, where $\vf$ is either $T$-periodic in $t$ or independent of $t$.
A perturbation $\xp = \vx - \xb$ satisfies $\dot{\vx}' = \mathrm{D}_{\vx}\vf(\xb(t),t)\,\xp + O(\abs{\xp}^2)$, whose linear part has the form \cref{eq:ltp}.
Its multipliers are the eigenvalues of the derivative of the period map (the Poincar\'e map) at $\xb(0)$, and they decide the stability of the orbit \cite{Hale-1980,Chicone-2006}.
When $\vf$ does not depend on $t$, the period is set by the orbit rather than imposed, and $\dot{\xb}$ is a periodic solution of the linearized equation, so one multiplier equals one.
Data taken near such an orbit therefore consist of a periodic base state plus the output of a linear periodic system, up to terms of second order in the distance from the orbit.
\Cref{sec:base-state} describes what the methods of \cref{sec:methods} return in that case.

%% file: sections/03_method.tex
\section{The Floquet decomposition and its ensemble form}
\label{sec:methods}

\subsection{DMD as a least-squares fit}
\label{sec:dmd}

DMD takes a sequence of vectors $\vb{q}_1,\dots,\vb{q}_K$ and looks for one constant matrix that advances each to the next.
Collect them as the columns of a matrix $\mV = [\,\vb{q}_1\ \ \vb{q}_2\ \cdots\ \vb{q}_K\,]$, write $\mV_{a:b}$ for its columns $a$ to $b$, and let
\begin{equation}
  \Vx = \begin{bmatrix} \vb{q}_1 & \vb{q}_2 & \cdots & \vb{q}_{K-1} \end{bmatrix},
  \qquad
  \Vy = \begin{bmatrix} \vb{q}_2 & \vb{q}_3 & \cdots & \vb{q}_{K} \end{bmatrix},
  \label{eq:XYgeneric}
\end{equation}
be its first and last $K-1$ columns; the requirement is $\Vy = \mA\Vx$, where $K$, the number of columns of $\mV$, is $N$ for standard DMD of a record of $N$ snapshots and is given for the other methods in \cref{tab:methods}.
For field data, whose vectors have many more entries than there are columns, many matrices satisfy this exactly, and DMD takes the one of smallest norm,
\begin{equation}
  \mA = \Vy\Vx^{\pinv},
  \label{eq:A-pinv}
\end{equation}
where $\Vx^{\pinv}$ is the Moore--Penrose pseudo-inverse of $\Vx$; when no matrix satisfies it exactly, the same formula gives the best fit in the least-squares sense.
The matrix $\mA$, with as many rows and columns as a snapshot has entries, is never formed.
Instead one computes the singular value decomposition $\Vx\approx\mU\mSigma\mW^{\herm}$, truncated to its $r$ leading terms: the orthonormal columns of $\mU$ are the directions along which the columns of $\Vx$ vary most (their proper orthogonal modes), and the singular values in $\mSigma$ measure by how much.
Then $\Vx^{\pinv} = \mW\mSigma^{-1}\mU^{\herm}$, and the action of $\mA$ within the span of $\mU$ is the $r\times r$ matrix
\begin{equation}
  \mS = \mU^{\herm}\mA\,\mU = \mU^{\herm}\, \Vy\, \mW\, \mSigma^{-1} .
  \label{eq:Stilde}
\end{equation}
The DMD eigenvalues are the eigenvalues $\lambda$ of $\mS$, and the DMD mode that belongs to $\lambda$ is $\mU\vz$, where $\mS\vz = \lambda\vz$ \cite{Schmid-2010,Tu-2014}; the cost is one singular value decomposition and one small eigenvalue problem.

Two features of \cref{eq:Stilde} recur below.
The rank $r$ is the number of independent directions that the model retains: the number of modes present for exact data from a linear system, and a choice for measured data (\cref{sec:algorithm}).
The factor $\mSigma^{-1}$ divides by the singular values, which amplifies the noise in directions along which the data scarcely vary; truncation keeps those directions out of the model.

\subsubsection{When DMD is exact}
\label{sec:dmd-exact}

Suppose that the vectors are produced by a constant matrix $\mF$ acting repeatedly on a starting vector $\vz_1$, and that each iterate is seen through a constant matrix $\mO$:
\begin{equation}
  \vb{q}_k = \mO\,\vz_k, \qquad \vz_{k+1} = \mF\vz_k, \qquad k = 1,\dots,K .
  \label{eq:krylov-data}
\end{equation}
Neither $\mF$ nor $\mO$ need be known, and $\mO$ need not be square.
Only the part of $\mF$ that the starting vector excites can leave a trace in the data.
The iterates $\vz_1,\vz_2 = \mF\vz_1,\vz_3 = \mF^2\vz_1,\dots$ span a subspace $\Kry$, the Krylov subspace of $\mF$ and $\vz_1$, of some dimension $r$, and $\mF$ maps $\Kry$ into itself.
If $\mO$ is one-to-one on $\Kry$, the data factor as
\begin{equation}
  \Vx = \mG\mR, \qquad \Vy = \mG\,\widehat{\mF}\,\mR ,
  \label{eq:factor}
\end{equation}
where the $r$ columns of $\mG$ are the images under $\mO$ of a basis of $\Kry$, the $r\times r$ matrix $\widehat{\mF}$ represents $\mF$ on $\Kry$ in that basis, and the columns of $\mR$ are the coordinates of the iterates.
The matrix $\mG$ has full column rank because $\mO$ loses nothing on $\Kry$, and $\mR$ has full row rank once it has at least $r$ columns, that is, once $K-1\ge r$.
Then $\Vx^{\pinv} = \mR^{\pinv}\mG^{\pinv}$, and \cref{eq:A-pinv} gives
\begin{equation}
  \mA = \mG\,\widehat{\mF}\,\mG^{\pinv} ,
  \label{eq:A-GFG-text}
\end{equation}
a matrix whose eigenvalues are those of $\widehat{\mF}$ together with zeros.
The conclusion (\cref{lem:krylov} of \cref{app:proofs}) is the following.
\begin{quote}
\emph{DMD of rank $r$ applied to the sequence \cref{eq:krylov-data} returns exactly the eigenvalues of $\mF$ that the starting vector excites, and the mode returned with an eigenvalue is $\mO\vw$, the image of the corresponding eigenvector $\vw$ of $\mF$, provided that $\mO$ is one-to-one on the excited subspace $\Kry$ and that there are at least $r$ column pairs.}
\end{quote}
The statement makes two requirements visible.
The data must be \emph{rich} enough, in that only the excited part of the spectrum can be recovered, and the measurement must be \emph{one-to-one} on what is excited.
Neither involves the number of entries of the vectors beyond its being at least $r$.
Every exactness result in this paper is an application of this statement; the methods differ only in what plays the part of $\mF$ and of $\mO$.

\subsubsection{One computation, four methods}
\label{sec:four-methods}

\begin{figure}
\centering
\newcommand{\cellborders}[3]{\foreach \i in {1,...,#2}{\foreach \j in {1,...,#3}{\draw[line width=0.3pt] (#1-\i-\j.north west) rectangle (#1-\i-\j.south east);}}}
\begin{tikzpicture}[font=\scriptsize,
  cell/.style={minimum width=6.3mm, minimum height=4.4mm, inner sep=0pt, outer sep=0pt, anchor=center, text height=1.2ex, text depth=0.45ex},
  ph0/.style={fill=white}, ph1/.style={fill=black!14}, ph2/.style={fill=black!36},
  grid/.style={matrix of math nodes, nodes={cell}, column sep=0pt, row sep=0pt, inner sep=0pt}]
  \matrix (rec) [grid] at (0,0) {
    |[ph0]| \snap_1 & |[ph1]| \snap_2 & |[ph2]| \snap_3 & |[ph0]| \snap_4 & |[ph1]| \snap_5 & |[ph2]| \snap_6 &
    |[ph0]| \snap_7 & |[ph1]| \snap_8 & |[ph2]| \snap_9 & |[ph0]| \snap_{10} & |[ph1]| \snap_{11} & |[ph2]| \snap_{12} \\};
  \cellborders{rec}{1}{12}
  \node[right=0.5mm of rec] {$\cdots$};
  \node[left=1.5mm of rec, font=\small] {(a)};
  \draw[decorate, decoration={brace, amplitude=3pt}] ([yshift=2pt]rec-1-1.north west) -- node[above=3pt] {one period, $p=3$} ([yshift=2pt]rec-1-3.north east);
  \draw[decorate, decoration={brace, amplitude=3pt, mirror}] ([yshift=-2pt]rec-1-1.south west) -- node[below=3pt] {$\dt$} ([yshift=-2pt]rec-1-1.south east);
  \node[font=\small] at (-3.35,-1.45) {columns one sample apart ($s=1$)};
  \node[font=\small] at (3.35,-1.45) {columns one period apart ($s=p$)};
  \matrix (dmd) [grid] at (-3.35,-2.45) {
    |[ph0]| \snap_1 & |[ph1]| \snap_2 & |[ph2]| \snap_3 & |[ph0]| \snap_4 & |[ph1]| \snap_5 \\};
  \cellborders{dmd}{1}{5}
  \node[right=0.5mm of dmd] {$\cdots$};
  \node[above=0.6mm of dmd, font=\small] {(b) DMD};
  \matrix (fd) [grid] at (3.35,-2.45) {
    |[ph0]| \snap_1 & |[ph0]| \snap_4 & |[ph0]| \snap_7 & |[ph0]| \snap_{10} \\};
  \cellborders{fd}{1}{4}
  \node[right=0.5mm of fd] {$\cdots$};
  \node[above=0.6mm of fd, font=\small] {(c) FD};
  \matrix (hk) [grid] at (-3.35,-4.55) {
    |[ph0]| \snap_1 & |[ph1]| \snap_2 & |[ph2]| \snap_3 & |[ph0]| \snap_4 & |[ph1]| \snap_5 \\
    |[ph1]| \snap_2 & |[ph2]| \snap_3 & |[ph0]| \snap_4 & |[ph1]| \snap_5 & |[ph2]| \snap_6 \\
    |[ph2]| \snap_3 & |[ph0]| \snap_4 & |[ph1]| \snap_5 & |[ph2]| \snap_6 & |[ph0]| \snap_7 \\};
  \cellborders{hk}{3}{5}
  \node[right=0.5mm of hk] {$\cdots$};
  \node[above=0.6mm of hk, font=\small] {(d) Hankel-DMD};
  \matrix (efd) [grid] at (3.35,-4.55) {
    |[ph0]| \snap_1 & |[ph0]| \snap_4 & |[ph0]| \snap_7 & |[ph0]| \snap_{10} \\
    |[ph1]| \snap_2 & |[ph1]| \snap_5 & |[ph1]| \snap_8 & |[ph1]| \snap_{11} \\
    |[ph2]| \snap_3 & |[ph2]| \snap_6 & |[ph2]| \snap_9 & |[ph2]| \snap_{12} \\};
  \cellborders{efd}{3}{4}
  \node[right=0.5mm of efd] {$\cdots$};
  \node[above=0.6mm of efd, font=\small] {(e) EFD};
  \node[rotate=90, font=\small] at (-6.3,-2.45) {$d=1$};
  \node[rotate=90, font=\small] at (-6.3,-4.55) {$d=p$};
\end{tikzpicture}
\caption{How the four methods of \cref{tab:methods} build their data matrices from one record, drawn for $p=3$ samples per period; the shading marks the phase within the period.
(a) The record of snapshots.
(b) Standard DMD uses the snapshots as they come.
(c) FD keeps one snapshot per period, all at the same phase.
(d) Hankel-DMD stacks $p$ consecutive snapshots and steps by one sample, so that the phases cycle along every row.
(e) EFD stacks the same $p$ snapshots and steps by one period: each row is the FD matrix of one phase.
In every case DMD fits the one matrix that advances each column to the next.}
\label{fig:matrices}
\end{figure}
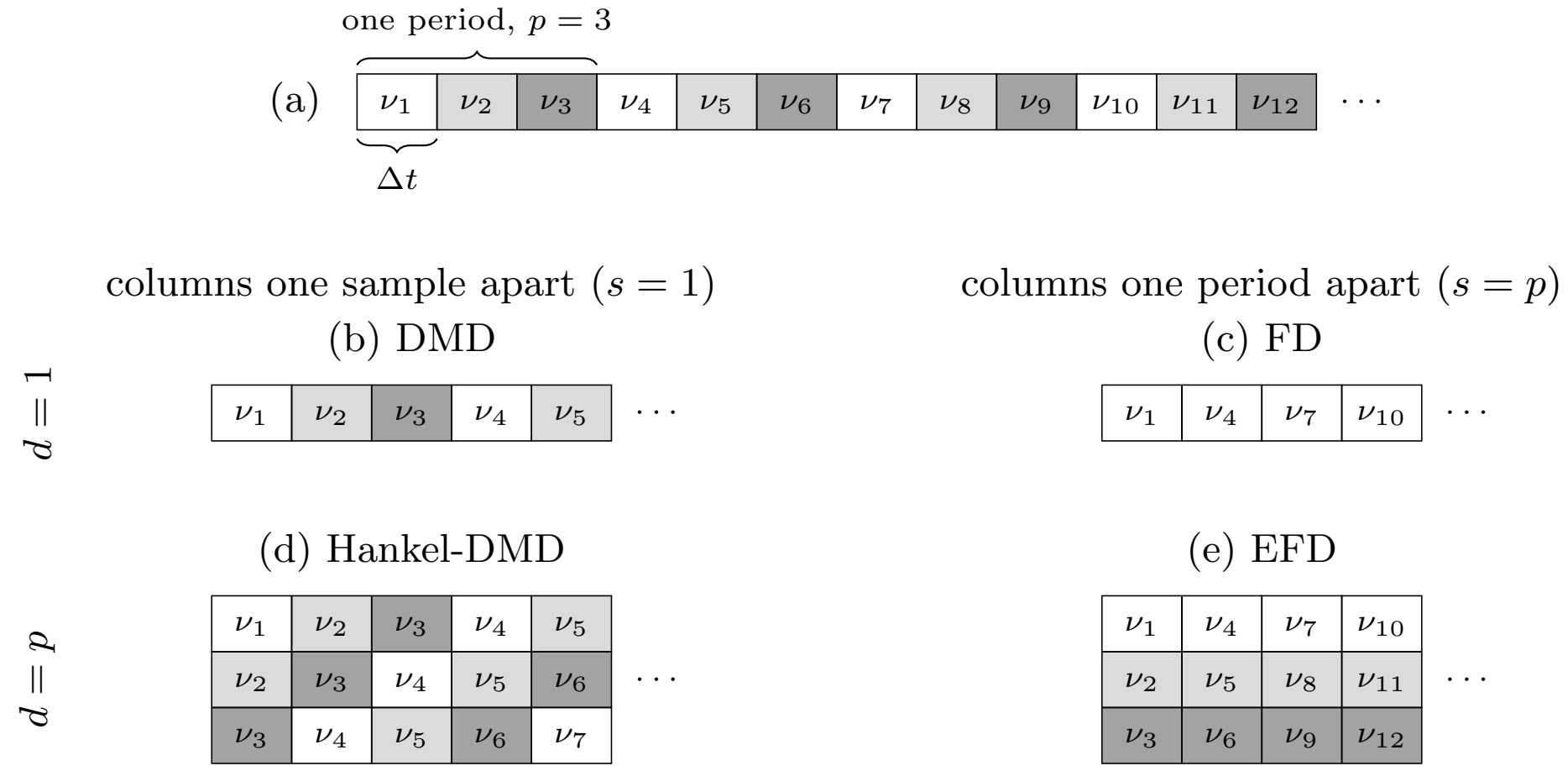

The vectors given to DMD need not be single snapshots.
For a window length $d\ge 1$ and a column stride $s\ge 1$, stack $d$ consecutive snapshots into the vector $\stk_k^{(d)}$ of \cref{eq:delay-vector} and take every $s$-th one:
\begin{equation}
  \begin{gathered}
  \Vx = \begin{bmatrix} \stk_1^{(d)} & \stk_{1+s}^{(d)} & \cdots & \stk_{1+(K-2)s}^{(d)} \end{bmatrix},
  \\
  \Vy = \begin{bmatrix} \stk_{1+s}^{(d)} & \stk_{1+2s}^{(d)} & \cdots & \stk_{1+(K-1)s}^{(d)} \end{bmatrix}.
  \end{gathered}
  \label{eq:XY}
\end{equation}
These are the first and last $K-1$ columns of a block-Hankel matrix with window $d$, stride $s$, and $K$ columns, the $k$-th of which is $\stk_{1+(k-1)s}^{(d)}$; $K$ is the number of windows that fit in the record when they are placed $s$ samples apart.
We write $\mathrm{DMD}(d,s)$ for the rank-$r$ DMD of this pair.
The window decides how much of the record each column sees, and the stride decides the time step of the matrix that is fitted.
\Cref{tab:methods} lists the four cases that concern us, and \cref{fig:matrices} shows how each builds its columns from the same record.

\begin{table}[tbhp]
\caption{The methods compared in this paper as DMD on a block-Hankel matrix with window $d$, column stride $s$, and $K$ columns.
$N$ is the number of snapshots in the record and $p$ the number of samples per period, so that $N/p$ is the number of periods; the counts for FD and EFD are for a record of whole periods.
Standard DMD is that of Schmid~\cite{Schmid-2010}, and Hankel-DMD that of Arbabi and Mezi\'c~\cite{Arbabi-2017} and Le~Clainche and Vega~\cite{LeClainche-2017}.}
\label{tab:methods}
\centering
\small
\begin{tabular}{@{}lcccl@{}}
\toprule
Method & $d$ & $s$ & $K$ & Matrix that is fitted \\ \midrule
Standard DMD & $1$ & $1$ & $N$ & one-sample map, if one exists \\
Hankel-DMD & $d$ & $1$ & $N-d+1$ & one-sample map of the stacked snapshots \\
FD & $1$ & $p$ & $N/p$ & monodromy matrix \\
EFD & $p$ & $p$ & $N/p$ & monodromy matrix \\ \bottomrule
\end{tabular}
\end{table}

\subsection{Floquet decomposition}
\label{sec:fd}

Schmid's observation \cite[\S\,2.6]{Schmid-2010} concerns a sampling rate tuned to the oscillation, but the sampling rate of a high-speed measurement is fixed by the hardware, ideally well above the frequency of interest, and the period is unknown when the data are taken.
The Floquet decomposition therefore keeps the full-rate record and \emph{subsamples} it, keeping every $p$-th snapshot, where $p$ is the number of samples per period \cref{eq:commensurate}: FD is $\mathrm{DMD}(1,p)$ applied to
\begin{equation}
  \begin{aligned}
  \mV^{(i)}_{1:K-1} &= \begin{bmatrix} \snap_i & \snap_{i+p} & \cdots & \snap_{i+(K-2)p} \end{bmatrix},
  \\
  \mV^{(i)}_{2:K} &= \begin{bmatrix} \snap_{i+p} & \snap_{i+2p} & \cdots & \snap_{i+(K-1)p} \end{bmatrix},
  \end{aligned}
  \label{eq:FD-data}
\end{equation}
for a starting phase $i\in\{1,\dots,p\}$ (\cref{fig:matrices}(c)), with $K$ the number of periods in the record.

\subsubsection{Why it works}

By \cref{eq:period-map} the columns of \cref{eq:FD-data} are
\begin{equation}
  \snap_{i+(k-1)p} = \mC_i\,\mM_i^{\,k-1}\,\vx_i , \qquad k = 1,2,\dots,K,
  \label{eq:FD-columns}
\end{equation}
successive iterates of the monodromy matrix at phase $i$, seen through the measurement at that phase.
This is the sequence \cref{eq:krylov-data} with $\mM_i$ in the part of $\mF$, $\mC_i$ in the part of $\mO$, and the state $\vx_i$ in the part of the starting vector $\vz_1$.
In the simplest case, in which the whole state is measured, \cref{eq:FD-columns} reads $\mV^{(i)}_{2:K} = \mM_i\mV^{(i)}_{1:K-1}$: the constant matrix that DMD looks for exists, and it is the monodromy matrix.

Let $\Kry_i$ be the subspace spanned by the period-sampled states,
\begin{equation}
  \Kry_i = \spn\{\vx_i,\ \mM_i\vx_i,\ \mM_i^{2}\vx_i,\ \dots\}, \qquad \nt = \dim\Kry_i .
  \label{eq:Ki}
\end{equation}
Its dimension $\nt$ is the number of multipliers that the initial state excites; it does not depend on $i$, since $\mM_i^{\,k}\vx_i = \mPhi(t_i,0)\,\mM^k\vx_0$ by \cref{eq:Mi-similar}, and it equals $n$ when the multipliers are distinct and the initial state $\vx_0$ has a non-zero component along every eigenvector of $\mM$.
The result of \cref{sec:dmd-exact} then gives the following (\cref{prop:fd}).
\begin{quote}
\emph{If the measurement matrix $\mC_i$ is one-to-one on $\Kry_i$ and the record contains more than $\nt$ periods, FD of rank $\nt$ returns exactly the $\nt$ Floquet multipliers that the initial state excites.
The mode returned with $\mu_l$ is $\mC_i\vw_l^{(i)} = \ee^{\rho_l t_i}\,\mC_i\,\vphi_l(t_i)$, the observed Floquet mode at phase $t_i$.}
\end{quote}

No assumption is made about how $\mA(t)$ varies within the period.
It need not vary slowly, and its values at different times need not commute; only the periodicity is used.
The price is the condition on $\mC_i$, which requires at least as many independent measurements at one instant as there are excited directions, $m\ge \nt$.
A single displacement signal from a second-order oscillator does not satisfy it: one number cannot determine a displacement and a velocity.

\subsubsection{Which multipliers are found, and at what frequencies}
\label{sec:leading}

With exact data FD returns every multiplier that the initial state excites; with noisy data and a truncation rank below $\nt$ it returns approximations to those that contribute most to the record, and since the contribution of $\mu_l$ to the $k$-th column is proportional to $\mu_l^{\,k}$, long records favor the multipliers of largest modulus, which decide stability.

An FD eigenvalue $\muh$ gives the growth rate $\re\rhoh = \ln\abs{\muh}/T$ without ambiguity, but the frequency only modulo $\omega$,
\begin{equation}
  \im\rhoh = \frac{\arg\muh}{T} + j\,\omega, \qquad j \in \ZZ ,
  \label{eq:unwrap}
\end{equation}
as in \cref{eq:exponent}: an aliasing with respect to the period, not the sampling interval, which is intrinsic to Floquet theory, and whose integer $j$ must be fixed from other information, a spectrum of the full-rate data or the harmonic content of the EFD mode (\cref{sec:efd-modes}).

\subsection{Ensemble Floquet decomposition}
\label{sec:efd}

The FD analyses started at different phases estimate the same multipliers from disjoint subsets of the record; run separately on noisy data they return slightly different eigenvalues, and their modes cannot then be paired or averaged with confidence.
The ensemble Floquet decomposition processes all phases in one eigenvalue problem instead.
It is $\mathrm{DMD}(p,p)$: each column is the stack of the $p$ consecutive snapshots of one period,
\begin{equation}
  \begin{gathered}
  \Vxt = \begin{bmatrix} \stk_1^{(p)} & \stk_{p+1}^{(p)} & \cdots & \stk_{(K-2)p+1}^{(p)} \end{bmatrix},
  \\
  \Vyt = \begin{bmatrix} \stk_{p+1}^{(p)} & \stk_{2p+1}^{(p)} & \cdots & \stk_{(K-1)p+1}^{(p)} \end{bmatrix},
  \\[4pt]
  \stk_{(k-1)p+1}^{(p)} = \begin{bmatrix} \snap_{(k-1)p+1} \\ \snap_{(k-1)p+2} \\ \vdots \\ \snap_{kp} \end{bmatrix}, \qquad k = 1,\dots,K .
  \end{gathered}
  \label{eq:EFD-data}
\end{equation}
Column $k$ is the $k$-th period of the record, so $K$ is again the number of periods, and the $c$-th block row of $\Vxt$ is the FD data matrix $\mV^{(c)}_{1:K-1}$ of \cref{eq:FD-data} for phase $c$ (\cref{fig:matrices}(e)): the ensemble in the name is the set of $p$ period-sampled sequences, one for each phase, drawn from a single trajectory, not an ensemble of trajectories or of initial conditions.

\subsubsection{Why it works, and what it requires}

By \cref{eq:h-equals-Ox,eq:period-map} the stacked columns are
\begin{equation}
  \stk_{(k-1)p+1}^{(p)} = \mObs_1^{(p)}\,\vx_{(k-1)p+1} = \mObs_1^{(p)}\,\mM_1^{\,k-1}\,\vx_1 , \qquad k = 1,\dots,K .
  \label{eq:EFD-columns}
\end{equation}
This is again the sequence \cref{eq:krylov-data}: iterates of the monodromy matrix, now seen through the tall constant matrix $\mObs_1^{(p)}$, whose block $c$ carries the state from the first phase of the period to phase $c$ and measures it there.
The matrix being iterated is the same as in FD.
What has changed is the measurement, and with it the requirement (\cref{prop:efd}).
\begin{quote}
\emph{If the system is observable over one period on $\Kry_1$, that is, if the excited state can be reconstructed from the measurements made during one period, and the record contains more than $\nt$ periods, then the stacked matrix $\Vxt$ has rank $\nt$, and EFD of rank $\nt$ returns exactly the $\nt$ Floquet multipliers that the initial state excites.
The mode returned with $\mu_l$ is $\vpsit_l = \mObs_1^{(p)}\vw_l^{(1)}$, whose $c$-th block is}
\begin{equation}
  \begin{gathered}
  \vpsit_l^{[c]} = \mC_c\,\mPhi(t_c,t_1)\,\vw_l^{(1)} = \mC_c\,\mPhi(t_c,0)\,\vw_l = \ee^{\rho_l t_c}\,\vpsi_l(t_c),
  \\
  \vpsi_l(t_c) = \mC_c\,\vphi_l(t_c), \qquad c = 1,\dots,p .
  \end{gathered}
  \label{eq:efd-mode-block}
\end{equation}
\end{quote}
Here $\vpsi_l$ is the \emph{observed Floquet mode}, the Floquet mode as the measurement sees it.
Three consequences follow.

First, stacking does not raise the rank of the data matrix: every block row of $\Vxt$ is a linear image, $\mC_c\mPhi(t_c,t_1)$, of the same sequence of states $\vx_1$, $\mM_1\vx_1$, $\mM_1^2\vx_1, \dots$, so the stacked matrix has the rank $\nt$ of that sequence however many phases are stacked.
What stacking changes is the hypothesis.
FD needs $\mC_i$ to be one-to-one, a condition on the measurement at one instant; EFD needs only that the excited state can be reconstructed from the measurements made during one period, which is the weaker and more natural requirement when the measurement is partial, because components of the state that are invisible to $\mC_c$ at one phase are carried into view by $\mPhi(t_c,t_1)$ at another, e.g., as the displacement of an oscillator, measured through one period, determines its velocity.
For a scalar measurement of the Mathieu equation, FD fails and EFD is exact (\cref{sec:verif-mathieu}).

Second, all of the data are used, and all phases share one set of multipliers by construction; the regression that determines $\mS$ has $pm$ rows instead of $m$.
We do not analyze the effect of noise here; \cref{sec:experiment} shows that it matters for measured data.

Third, the mode \cref{eq:efd-mode-block} is the observed Floquet mode resolved at the $p$ phases of the period (\cref{sec:efd-modes}).
The stack may also contain any subset of the phases, for example every second sample of the period, with $\mObs_1^{(p)}$ replaced by the corresponding block rows; FD is the case of a single phase.

\subsection{Phase-resolved and averaged modes}
\label{sec:efd-modes}

By \cref{eq:efd-mode-block} the blocks of an EFD mode differ from samples of the $T$-periodic function $\vpsi_l$ by the known factor $\ee^{\rho_l t_c}$.
Removing that factor aligns the blocks: the vectors $\ee^{-\rho_l t_c}\vpsit_l^{[c]} = \vpsi_l(t_c)$, $c=1,\dots,p$, are the time-periodic mode itself, one frame for each phase of the period.
A discrete Fourier transform across the aligned blocks gives the harmonics of the mode.
For every integer $j$,
\begin{equation}
  \frac{1}{p}\sum_{c=1}^{p} \ee^{-(\rho_l + \ii j\omega)t_c}\, \vpsit_l^{[c]}
  = \frac{1}{p}\sum_{c=1}^{p} \ee^{-2\pi\ii\, jc/p}\,\vpsi_l(t_c)
  =: \vpsih_{l,j}
  \label{eq:harmonic}
\end{equation}
is the $j$-th discrete Fourier coefficient, or harmonic, of the observed Floquet mode, since $\omega t_c = 2\pi c/p$ (\cref{prop:harmonics}).
Harmonic zero, $\vpsih_{l,0}$, is the plain average of the aligned blocks, the mean of the mode over the period.

Which function is called ``the mode'' depends on the branch of the exponent: changing $\rho_l$ to $\rho_l+\ii j\omega$ turns the mean into the $j$-th harmonic (\cref{sec:floquet-theory}).
The natural branch is the one for which the mean carries the most of the mode, and \cref{eq:harmonic} supplies it, since the transform returns all $\vpsih_{l,j}$ at once and the $j$ of largest norm identifies the dominant frequency $\im\rho_l + j\omega$, which fixes the integer left open in \cref{eq:unwrap} from the EFD output alone.
\Cref{sec:relations} shows that $\vpsih_{l,j}$ is also the Koopman mode of the lattice eigenvalue $\rho_l+\ii j\omega$ and, for any window long enough to distinguish the roots of $\mu_l$, the leading block of the mode that Hankel-DMD returns at the corresponding root.

For measured data the average also serves a statistical purpose: the $p$ aligned blocks are $p$ estimates of the same field, each computed from a different subset of the samples, and averaging them suppresses the part of each that is not coherent from phase to phase.
For real data the modes occur in complex-conjugate pairs, and the real and imaginary parts of the average are two real fields in quadrature, which together make up the mode.

\subsection{Periodic base states}
\label{sec:base-state}

Measured data are seldom a perturbation alone; they contain the base flow as well.
Suppose that $\snap_k = \mC_k(\xb_k + \xp_k)$, where $\xb_{k+p} = \xb_k$ is a periodic base state and $\xp_k$ satisfies \cref{eq:sampled}, as for small perturbations of a periodic orbit (\cref{sec:nonlinear-background}).
Then the stacked columns are $\stk_{(k-1)p+1}^{(p)} = \bar{\stk} + \mObs_1^{(p)}\mM_1^{\,k-1}\xp_1$, with $\bar{\stk}$ the stacked base state, the same in every column.
A constant is an iterate of the number one, so this is the sequence \cref{eq:krylov-data} for the matrix $\blkdiag(1,\mM_1)$ started from $(1,\xp_1)$.
FD and EFD therefore return the eigenvalue $1$, with the base state $\bar{\stk}$ as its mode, in addition to the multipliers excited by the perturbation, provided the augmented measurement $[\,\bar{\stk}\ \ \mObs_1^{(p)}\,]$ is one-to-one on the augmented excited subspace.

The entire periodic motion, whatever its harmonic content, thus occupies one dimension of the decomposition.
With a stride of one sample, by contrast, DMD represents a periodic signal by its discrete Fourier components, one eigenvalue on the unit circle for each harmonic present \cite{Chen-2012}.
For an autonomous system the trivial multiplier of the orbit is also one; a perturbation along the orbit shifts the phase of the base state and is absorbed in the same mode.

\subsection{Computation}
\label{sec:algorithm}

\subsubsection{The algorithm}

Given snapshots $\snap_1,\dots,\snap_N\in\CC^m$ at interval $\dt$, the number $p$ of samples per period and a rank $r$, EFD consists of the following steps; FD is the same with a single block.
\begin{enumerate}
\item[(i)] Set $K = \lfloor N/p\rfloor$, the number of whole periods in the record, form the stacked columns $\stk_{(k-1)p+1}^{(p)} = [\,\snap_{(k-1)p+1};\,\snap_{(k-1)p+2};\,\dots;\,\snap_{kp}\,]$ for $k=1,\dots,K$, and the matrices $\Vxt$ and $\Vyt$ of \cref{eq:EFD-data}.
\item[(ii)] Compute the rank-$r$ singular value decomposition $\Vxt\approx\mU\mSigma\mW^{\herm}$.
\item[(iii)] Form $\mS = \mU^{\herm}\Vyt\mW\mSigma^{-1}$ and solve $\mS\vz_l = \muh_l\vz_l$; the multipliers are $\muh_l$ and the stacked modes $\vpsit_l = \mU\vz_l$.
\item[(iv)] Set $\rhoh_l = \ln(\muh_l)/(p\,\dt)$ on the principal branch; $\re\rhoh_l$ is the growth rate and $\im\rhoh_l$ the frequency modulo $\omega$.
\item[(v)] Split $\vpsit_l$ into its $p$ blocks $\vpsit_l^{[c]}\in\CC^m$, $c = 1,\dots,p$, and align them, $\vpsi_l(t_c) = \ee^{-\rhoh_l c\,\dt}\vpsit_l^{[c]}$: this is the phase-resolved mode.
\item[(vi)] Take the discrete Fourier transform across the blocks, the harmonics $\vpsih_{l,j} = p^{-1}\sum_c \ee^{-2\pi\ii\,jc/p}\vpsi_l(t_c)$, $j=0,\dots,p-1$, and let $j^\star$ be the harmonic of largest norm.
Report the frequency $\im\rhoh_l + 2\pi j^\star/(p\,\dt)$, aliased to the band resolved by $\dt$, and the averaged mode $\vpsih_{l,j^\star}$.
\end{enumerate}

\subsubsection{Cost, and the method of snapshots}

The cost is that of one DMD of a matrix with $pm$ rows and $K-1$ columns, $p$ times fewer columns than Hankel-DMD has for the same window.
For field data the stacked matrix is tall, $pm\gg K$, and need not be formed: the singular vectors $\mW$ and the singular values follow from the matrix of inner products of its columns, $\Vx^{\herm}\Vx = \mW\mSigma^2\mW^{\herm}$, the method of snapshots of Sirovich~\cite{Sirovich-1987}, and then
\begin{equation}
  \mS = \mSigma^{-1}\mW^{\herm}\bigl(\Vx^{\herm}\Vy\bigr)\mW\mSigma^{-1}, \qquad \mU\vz_l = \Vx\,\mW\mSigma^{-1}\vz_l ,
  \label{eq:gram-S}
\end{equation}
because $\mU = \Vx\mW\mSigma^{-1}$.
The inner product of two stacked columns is the sum of the inner products of their blocks,
\begin{equation}
  \bigl(\stk_a^{(d)}\bigr)^{\herm}\stk_b^{(d)} = \sum_{c=1}^{d} \snap_{a+c-1}^{\herm}\,\snap_{b+c-1} ,
  \label{eq:gram}
\end{equation}
so $\Vx^{\herm}\Vx$ and $\Vx^{\herm}\Vy$, for any window and stride, are sums of shifted sub-blocks of the one $N\times N$ matrix of inner products of the raw snapshots, which is computed once; afterwards a change of method, window, stride, period, or rank costs only a small eigenvalue problem, which makes the sensitivity studies of \cref{sec:exp-sensitivity} cheap.

Working from inner products squares the singular values, so those below about $10^{-8}$ of the largest are lost in double precision; this is immaterial for measured data but not for tests at the level of round-off, for which \cref{sec:verification} uses the singular value decomposition of $\Vx$ itself.

\subsubsection{The rank}

For exact data from a linear periodic system the rank is not a choice: the singular values of $\Vx$ fall to round-off after the $\nt$-th, and $r=\nt$.
For measured data they decay without a gap, and $r$ is a regularization parameter: too small a rank leaves out the directions that carry the modes of interest, and a larger one admits more of the record, noise included, which moves the moduli but hardly the frequencies (\cref{sec:exp-sensitivity}).
We know of no rule that removes the choice, and recommend reporting the dependence on $r$.

\subsubsection{Scaling of the data}

DMD is a least-squares fit in the Euclidean norm of the columns, so multiplying a group of rows by a factor changes how much attention the fit pays to them.
For exact data from a linear system this changes no eigenvalue, since it replaces $\mC_k$ by $\mD\mC_k$ with $\mD$ an invertible diagonal matrix; for measured data it decides which measurements define the leading singular vectors, and so what survives the truncation.
The relative weights of the kinds of data in a composite snapshot are therefore a choice that must be stated and varied (\cref{sec:exp-sensitivity}); weighting each row by the confidence in it, as particle image velocimetry estimates point by point, with weights fixed in time, is a natural refinement that we leave to others.

\subsubsection{Bias}

The least-squares fit \cref{eq:A-pinv} treats $\Vx$ as exact and assigns all of the misfit to $\Vy$.
When both contain noise, the fitted matrix is biased toward zero, and its eigenvalues toward the origin \cite{Dawson-2016,Hemati-2017}.
Dynamics that the linear periodic model cannot represent enter the regression as noise does (\cref{sec:verif-nonlinear}).
The modulus of a multiplier estimated from measured data is therefore biased low by an unknown amount, a point that governs the interpretation of the results in \cref{sec:experiment}.

%% file: sections/04_relations.tex
\section{Relation to Hankel-DMD, the Koopman lattice, and other methods}
\label{sec:relations}

\begin{figure}
\centering
\begin{tikzpicture}[>=stealth, font=\small]
  \begin{scope}[shift={(0,0)}]
    \draw[->] (-1.9,0) -- (0.6,0) node[below] {$\re\rho$};
    \draw[->] (0,-2.2) -- (0,2.4) node[left] {$\im\rho$};
    \foreach \j in {-3,...,3} {
      \fill (-1.0,{0.6*(\j+0.278)}) circle (1.7pt);
      \draw (-1.0,{0.6*(\j-0.278)}) circle (1.7pt);
    }
    \draw[<->] (-1.45,{0.6*0.278}) -- node[left] {$\omega$} (-1.45,{0.6*1.278});
    \node at (-0.65,-2.75) {(a) lattice $\rho_l+\ii j\omega$};
  \end{scope}
  \draw[->, thick] (0.8,0.6) -- node[above] {sample} node[below] {at $\dt$} (2.1,0.6);
  \begin{scope}[shift={(3.8,0)}]
    \draw[->] (-1.55,0) -- (1.55,0);
    \draw[->] (0,-1.55) -- (0,1.55);
    \draw[gray] (0,0) circle (1.3);
    \foreach \j in {0,...,7} {
      \fill ({1.3*0.9725*cos(12.5+45*\j)},{1.3*0.9725*sin(12.5+45*\j)}) circle (1.5pt);
      \draw[fill=white] ({1.3*0.9725*cos(-12.5+45*\j)},{1.3*0.9725*sin(-12.5+45*\j)}) circle (1.5pt);
    }
    \node at (0,-2.75) {(b) comb $\beta_{l,j}=\mu_l^{1/p}\zeta^{\,j}$};
  \end{scope}
  \draw[->, thick] (5.55,0.6) -- node[above] {sample at $T$} node[below] {($p$-th power)} (7.2,0.6);
  \begin{scope}[shift={(8.9,0)}]
    \draw[->] (-1.55,0) -- (1.55,0);
    \draw[->] (0,-1.55) -- (0,1.55);
    \draw[gray] (0,0) circle (1.3);
    \fill ({1.3*0.8*cos(100)},{1.3*0.8*sin(100)}) circle (1.7pt);
    \draw ({1.3*0.8*cos(-100)},{1.3*0.8*sin(-100)}) circle (1.7pt);
    \node at (0,-2.75) {(c) multipliers $\mu_l$};
  \end{scope}
\end{tikzpicture}
\caption{The three levels at which the spectrum of a time-periodic system appears, drawn for a complex-conjugate pair of multipliers (filled and open symbols) and $p=8$ samples per period.
(a) In continuous time each Floquet mode is a carrier with sidebands: the exponents $\rho_l+\ii j\omega$, the Koopman lattice of Mezi\'c and Surana~\cite{Mezic-2016}.
(b) Sampling at $\dt$ aliases the lattice onto a comb of $p$-th roots of the multipliers, the eigenvalues of Hankel-DMD (\cref{sec:hankel}); the gray circle is the unit circle.
The teeth lie on a circle of radius $\abs{\mu_l}^{1/p}$, nearer the unit circle than the multiplier itself, which is why Hankel-DMD eigenvalues are compared with multipliers after raising them to the power $p$; the figure is drawn for $\abs{\mu_l} = 0.8$.
(c) Sampling at the period collapses each comb onto its Floquet multiplier, the eigenvalues of FD and EFD (\cref{sec:collapse}).}
\label{fig:three-levels}
\end{figure}
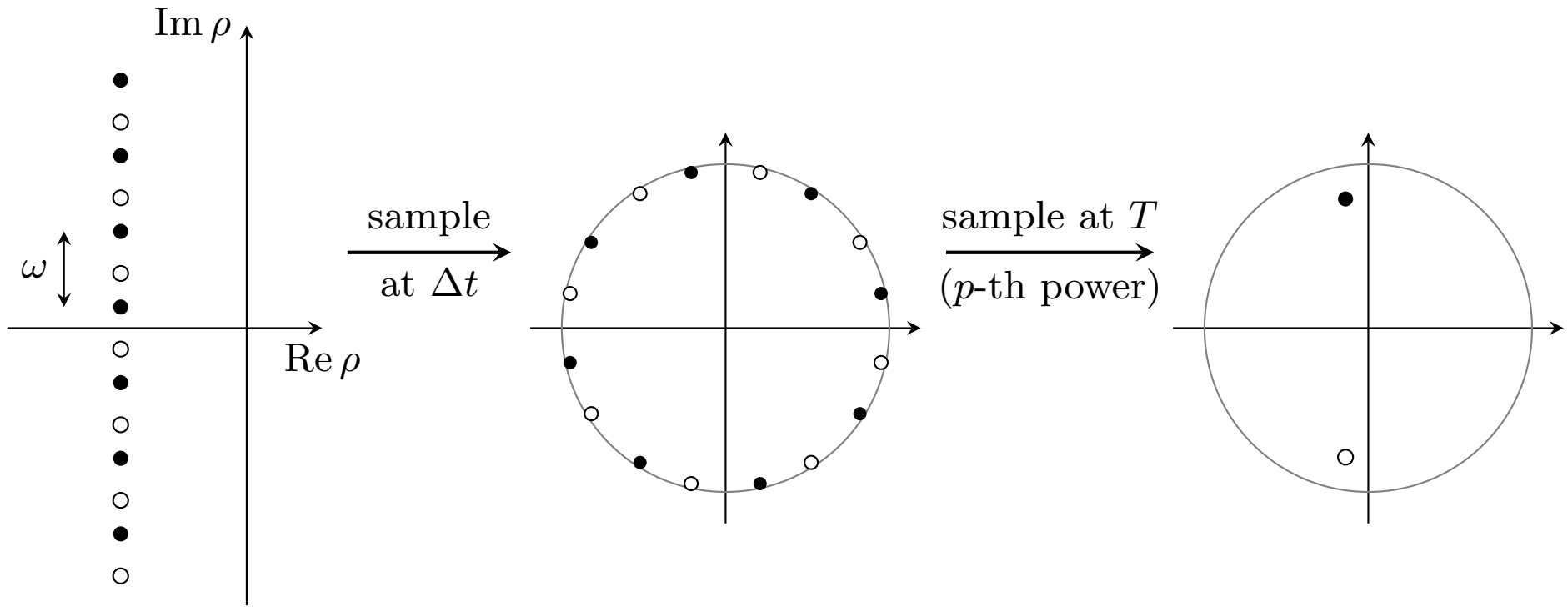

EFD and Hankel-DMD are built on the same stacked snapshots (\cref{tab:methods}, \cref{fig:matrices}(d) and (e)); with the window $d=p$, their data matrices differ in the stride between columns, one period against one sample.
This section determines what the stride-one computation returns for a linear periodic system, and how its output is related to that of FD and EFD and to the Koopman spectrum of the system.
\Cref{fig:three-levels} summarizes the answer.

\subsection{A Floquet mode at the sampling rate: carrier and sidebands}
\label{sec:sidebands}

Assume that $\mM$ is diagonalizable, with eigenpairs $(\mu_l,\vw_l)$ and $\vx_0 = \sum_la_l\vw_l$, and fix an exponent $\rho_l$ for each multiplier.
By the Floquet expansion \cref{eq:floquet-expansion} and $\snap_k = \mC_k\vx_k$, the snapshots are
\begin{equation}
  \snap_k = \sum_{l=1}^{n} a_l\,\ee^{\rho_l t_k}\,\vpsi_l(t_k), \qquad \vpsi_l(t_k) = \mC_k\,\vphi_l(t_k) .
  \label{eq:y-floquet}
\end{equation}
The observed Floquet mode $\vpsi_l$ repeats every $p$ samples, so it is the sum of its $p$ discrete harmonics \cref{eq:harmonic},
\begin{equation}
  \vpsi_l(t_k) = \sum_{j=0}^{p-1} \vpsih_{l,j}\,\ee^{2\pi\ii\,jk/p} .
  \label{eq:psi-dft}
\end{equation}
Substituting, and writing
\begin{equation}
  \beta_l = \ee^{\rho_l\dt}, \qquad \zeta = \ee^{2\pi\ii/p}, \qquad \beta_{l,j} = \beta_l\,\zeta^{\,j} = \ee^{(\rho_l+\ii j\omega)\dt}, \quad j = 0,\dots,p-1 ,
  \label{eq:teeth}
\end{equation}
we obtain
\begin{equation}
  \snap_k = \sum_{l=1}^{n}\sum_{j=0}^{p-1} a_l\,\vpsih_{l,j}\,\beta_{l,j}^{\,k} .
  \label{eq:comb-expansion}
\end{equation}

This is the carrier-and-sidebands picture of \cref{sec:floquet-theory} in sampled form: each term of \cref{eq:comb-expansion} is a pure exponential in $k$ with a fixed shape $\vpsih_{l,j}$, as a time-invariant linear system would produce, so that, seen at the sampling rate, a time-periodic system with $n$ Floquet modes is indistinguishable from a time-invariant one with as many as $np$ modes, those of one Floquet mode sharing the growth rate $\re\rho_l$ at frequencies spaced by the fundamental.

The $p$ numbers $\beta_{l,j}$ are the $p$ distinct $p$-th roots of the multiplier,
\begin{equation}
  \beta_{l,j}^{\,p} = \mu_l .
  \label{eq:tooth-power}
\end{equation}
They lie on a circle of radius $\abs{\mu_l}^{1/p}$, equally spaced in angle, a comb with $p$ teeth (\cref{fig:three-levels}(b)), and tooth $j$ of multiplier $l$ is present in the data exactly when the observed Floquet mode has a non-zero $j$-th harmonic.
Let
\begin{equation}
  \Jset = \bigl\{(l,j) : a_l\,\vpsih_{l,j}\ne \vb{0}\bigr\}, \qquad r = \abs{\Jset}\le np ,
  \label{eq:Jset}
\end{equation}
be the set of teeth present and their number.

\subsection{What Hankel-DMD returns}
\label{sec:hankel}

A stacked snapshot inherits the expansion.
Block $c$ of $\stk_k^{(d)}$ is $\snap_{k+c-1}$, so by \cref{eq:comb-expansion}
\begin{equation}
  \stk_k^{(d)} = \sum_{(l,j)\in\Jset} a_l\,\beta_{l,j}^{\,k}\; \vb{v}_d(\beta_{l,j})\otimes\vpsih_{l,j}, \qquad \vb{v}_d(\beta) = (1,\beta,\dots,\beta^{d-1})^{\trans},
  \label{eq:h-comb}
\end{equation}
where $\vb{v}_d(\beta)\otimes\vpsih$ denotes the stack of the $d$ blocks $\vpsih,\beta\vpsih,\dots,\beta^{d-1}\vpsih$.
In matrix form, $\stk_k^{(d)} = \mG\,\mD^k\vone$, with $\mD$ the diagonal matrix of the $r$ teeth, $\vone$ the vector of ones, and the columns of $\mG$ the vectors $a_l\,\vb{v}_d(\beta_{l,j})\otimes\vpsih_{l,j}$.
This is once more the sequence \cref{eq:krylov-data}: iterates of a constant matrix, here a diagonal one, seen through a constant matrix.
The result of \cref{sec:dmd-exact} applies as soon as the columns of $\mG$ are linearly independent, which a long enough window ensures (\cref{thm:hankel}).
\begin{quote}
\emph{Let the Floquet multipliers be distinct and the sampling commensurate.
If the window satisfies $d\ge r$ and the record supplies at least $r$ column pairs, the stride-one stacked matrix has rank $r$, and Hankel-DMD of rank $r$ returns exactly the teeth $\beta_{l,j}$, $(l,j)\in\Jset$.
Each eigenvalue is a $p$-th root of a Floquet multiplier, and the number of teeth returned for a multiplier equals the number of non-zero harmonics of its observed Floquet mode.
The mode returned with $\beta_{l,j}$ is $\vb{v}_d(\beta_{l,j})\otimes\vpsih_{l,j}$, whose leading block is the $j$-th harmonic $\vpsih_{l,j}$ of the observed Floquet mode.}
\end{quote}
We comment on the hypotheses and on what the result does and does not say.

\emph{The comb reflects harmonic content.}
A time-invariant system observed through a constant matrix has one tooth per eigenvalue, and the result reduces to the familiar statement that DMD recovers $\ee^{\lambda\dt}$; the Floquet modes of the Mathieu equation contain every harmonic, with amplitudes that decay geometrically for small $\varepsilon$, so all $np$ teeth are present in exact arithmetic but only the first few are above any practical noise floor (\cref{sec:verif-comb}).

\emph{Rank and window.}
The window must be long enough to distinguish the teeth that are present: $d\ge r$ is sufficient for every number $m$ of observables, $md\ge r$ is necessary, and for field data, where $m\gg r$, a short window can suffice.
On the measurements of \cref{sec:experiment}, however, the window had to span about one analysis period, the time over which the measurements determine enough of the state (\cref{sec:exp-sensitivity}).

\emph{Distinct multipliers.}
The assumption serves the explicit description by harmonics; without it every Hankel-DMD eigenvalue is still a $p$-th root of a multiplier, once the rank of the stacked matrix has stopped growing with $d$ and $K$ (\cref{app:hankel}).

\emph{Assembling the Floquet mode.}
The leading block of the Hankel-DMD mode at one tooth is the corresponding harmonic, and when harmonic zero dominates, as it does in the data of \cref{sec:experiment}, it is the phase-averaged EFD mode (\cref{sec:collapse}).
The phase-resolved mode, however, is the sum of the modes over all the teeth of the multiplier, \cref{eq:mode-sum}, and assembling it requires the teeth to be recognized as one family and combined with consistent phases; with measured data the $p$-th powers of the teeth of one multiplier do not coincide, and weak teeth are lost to the truncation.
EFD returns the family as one eigenvalue whose stacked mode is the phase-resolved mode.

\emph{Finite data, no ergodicity.}
The convergence theory of Hankel-DMD \cite{Arbabi-2017} concerns ergodic systems in the limit of infinite record length; the present result is exact for a finite record and applies to growing and decaying solutions, which is the situation of interest in a stability analysis.

\emph{Truncation.}
When the harmonic amplitudes decay, Hankel-DMD is used with a truncation rank $\tilde{r} < r$, and the teeth with the largest amplitudes are recovered approximately, with an error governed by the size of the discarded harmonics (\cref{sec:verif-comb}); FD and EFD do not face this trade-off, because they require rank $\nt$ whatever the harmonic content.

\subsection{Sampling at the period collapses the comb}
\label{sec:collapse}

By \cref{eq:tooth-power} the $p$-th power sends every tooth to its multiplier.
Taking the columns one period apart performs this collapse on the data.
The $k$-th column then starts at sample $(k-1)p+1$, and since $\beta_{l,j}^{\,(k-1)p+1} = \mu_l^{\,k-1}\beta_{l,j}$, \cref{eq:h-comb} becomes
\begin{equation}
  \stk_{(k-1)p+1}^{(d)} = \sum_{l} a_l\,\mu_l^{\,k-1}\,\vpsit_l , \qquad
  \vpsit_l = \sum_{j:(l,j)\in\Jset} \beta_{l,j}\,\vb{v}_d(\beta_{l,j})\otimes\vpsih_{l,j} .
  \label{eq:mode-sum}
\end{equation}
The harmonics of one Floquet mode, separate exponentials at stride one, advance together at stride $p$ and merge into the single vector $\vpsit_l$, the factor $\beta_{l,j}$ being the advance from the initial time to the first sample.
For $d=p$ this vector is the EFD mode of \cref{eq:efd-mode-block}: its block $c$, $\sum_j\beta_{l,j}^{\,c}\vpsih_{l,j} = \ee^{\rho_lt_c}\vpsi_l(t_c)$, is the inverse discrete Fourier transform of the harmonics.

The same statement holds for the fitted operators (\cref{cor:collapse}): let $\mA_1 = \Vy\Vx^{\pinv}$ be the stride-one operator and $\mA_p = \Vyt\Vxt^{\pinv}$ the operator of the same stacked snapshots taken with stride $p$, as in \cref{eq:EFD-data}.
Under the conditions of \cref{sec:hankel},
\begin{equation}
  \mA_p = \mA_1^{\,p}\,\mPi_p ,
  \label{eq:collapse-identity}
\end{equation}
where $\mPi_p$ is the orthogonal projector onto the range of the stride-$p$ data: the EFD operator is the restriction of the $p$-th power of the Hankel-DMD operator, whose non-zero eigenvalues are the excited multipliers, each repeated once for every one of its teeth, while those of $\mA_p$ are the same multipliers, each once; the mode of $\mA_p$ for $\mu_l$ is the sum \cref{eq:mode-sum} of the Hankel-DMD modes over the comb of $\mu_l$.

The stride thus decides how the same information is packaged.
Hankel-DMD distributes each Floquet mode over as many eigenvalues as the mode has harmonics, all of which must be resolved and then recognized as one family, whereas EFD returns the family as a single multiplier whose mode contains the harmonics as its dependence on phase; the rank required falls from the number of excited harmonics to the number of excited multipliers, and the eigenvalue returned is the quantity by which Floquet theory decides stability.
On the measurements of \cref{sec:experiment}, far from the idealized setting, the two methods still return the same two leading pairs, with matching modes (\cref{sec:exp-results}).

\subsection{The lifted and cyclic reformulations}
\label{sec:cyclic}

The identities beneath these results are not new.
It has long been known in the theory of periodic control systems that a discrete-time system with period $p$ can be rewritten as a time-invariant one in two ways \cite{Meyer-1975,Bittanti-2009}.
The \emph{lifted} reformulation samples the state once per period and collects the $p$ outputs of the period into one vector: it is the system $\vx_{kp+1} = \mM_1\vx_{(k-1)p+1}$, $\stk_{(k-1)p+1}^{(p)} = \mObs_1^{(p)}\vx_{(k-1)p+1}$ of \cref{eq:EFD-columns}.
The \emph{cyclic} reformulation keeps the original time step and enlarges the state instead, to $p$ copies of the state space, one for each phase, around which the state is passed in a cycle.
Its matrix $\mF$, of order $np$, has the teeth $\beta_{l,j}$ as its eigenvalues and satisfies $\mF^p = \blkdiag(\mM_1,\mM_2,\dots,\mM_p)$ (\cref{lem:cyclic}).
In these terms EFD is DMD applied to the output of the lifted system and Hankel-DMD is DMD applied to the output of the cyclic system, and \cref{eq:collapse-identity} is the counterpart for the fitted operators of the identity for $\mF^p$.
The contribution of this section is not these identities but what they imply about the two data-driven computations: which eigenvalues each returns from a single trajectory, under what conditions, and how the modes are related.

\subsection{Standard DMD}
\label{sec:standard-dmd}

The case $d=1$ of \cref{sec:hankel} is standard DMD, which returns the teeth if and only if the $r$ harmonics $a_l\vpsih_{l,j}$ are linearly independent vectors of $\CC^m$ and the record supplies at least $r$ column pairs (\cref{cor:standard}), and so requires $m\ge r$.
The condition separates two situations that are often not distinguished.
For a low-dimensional periodic system observed in full, $m=n$, it fails as soon as one Floquet mode has more than one harmonic: no constant matrix maps each sample to the next, and the eigenvalues of the least-squares compromise are not roots of the multipliers, misplacing them by several percent or more for the Mathieu equation (\cref{sec:verif-mathieu}).
For high-dimensional field data with a few active Floquet modes whose harmonics have linearly independent spatial structure the condition can hold, and standard DMD then returns each Floquet mode as the comb of its sidebands, to which the remarks of \cref{sec:collapse} apply.

\subsection{Other strides, inexact periods, and incommensurate sampling}
\label{sec:strides}

The argument of \cref{sec:collapse} applies to any stride $s$, the eigenvalues being $\beta_{l,j}^{\,s} = \ee^{(\rho_l+\ii j\omega)s\dt}$; if $s\dt = (1+\delta)T$, with $\delta$ the relative error in the analysis period, tooth $(l,j)$ is sent to
\begin{equation}
  \beta_{l,j}^{\,s} = \mu_l^{\,1+\delta}\,\ee^{2\pi\ii\,j\delta}
  \label{eq:period-error}
\end{equation}
instead of to $\mu_l$: the harmonics of a Floquet mode are fanned out by the angle $2\pi j\delta$ rather than superposed.
If the window, the rank, and the length of the record suffice to resolve the fanned-out teeth, each is recovered exactly, and the growth rates and unwrapped frequencies, $\ln(\beta_{l,j}^{\,s})/(s\dt) = \rho_l+\ii j\omega$, do not depend on the analysis period, although more eigenvalues are needed to represent each mode; if instead the rank is the minimal $\nt$, the fanned-out teeth must be fitted by one eigenvalue, and the result is in error (\cref{sec:period}).

The same reasoning covers a period that is not an integer number of samples: for any $\dt$ the Fourier series of $\vpsi_l(t) = \mC(t)\vphi_l(t)$ gives $\snap_k = \sum_la_l\sum_{j}\vpsih^{\infty}_{l,j}\,\ee^{(\rho_l+\ii j\omega)k\dt}$, the sum running over all integers $j$, to which the argument of \cref{sec:hankel} applies whenever finitely many harmonics are present; commensurate sampling is needed for the exact collapse at stride $p$, not for the comb.

\subsection{The Koopman lattice}
\label{sec:lattice}

Mezi\'c and Surana~\cite{Mezic-2016} established the Koopman-operator description of time-periodic systems: for the linear system \cref{eq:ltp}, suspended as the autonomous system $\dot{\vx} = \mA(\tau)\vx$, $\dot\tau = 1$, the Koopman eigenvalues that enter the expansion of the state are
\begin{equation}
  \rho_l + \ii j\omega , \qquad l = 1,\dots,n, \quad j \in \ZZ ,
  \label{eq:lattice}
\end{equation}
a lattice in which every Floquet exponent is repeated at all harmonics of the fundamental frequency, with the Fourier coefficients of the Floquet modes as the Koopman modes of the state \cite[equations~(29)--(36) and Theorem~4.1]{Mezic-2016} (they denote the exponents by $\mu_k$); they extend the description to nonlinear periodic systems near a stable solution and to quasi-periodic time dependence.

The results of this section are the sampled-data form of that lattice at two sampling intervals (\cref{fig:three-levels}).
Sampling at $\dt = T/p$ maps the lattice point $\rho_l+\ii j\omega$ to $\beta_{l,j}$, and since harmonics $j$ and $j+p$ coincide, the lattice is aliased onto the comb of \cref{sec:sidebands}, with the aliased Koopman mode $\vpsih_{l,j}$ at each tooth.
Sampling at $T$ maps every lattice point of exponent $l$ to $\mu_l$, which removes the degeneracy that the suspension introduces and leaves the spectrum of the monodromy matrix, estimated by FD and EFD, with the harmonics carried by the phase dependence of the EFD mode (\cref{sec:efd-modes}): in Koopman terms, FD is the DMD of the stroboscopic map.

Mezi\'c and Surana also sketch a computational procedure, extended or kernel DMD applied phase by phase to ensembles of trajectories started at controlled phases, and leave its validation to future work.
A single measured time series cannot supply such ensembles; FD and EFD address that case, with one trajectory, one regression in which all phases share one set of multipliers, and exactness under stated conditions.

\subsection{Other methods for time-varying and periodic systems}
\label{sec:other-methods}

\emph{Koopman operator families for non-autonomous systems.}
Ma\'ce\v{s}i\'c et al.~\cite{Macesic-2018} compute the time-dependent Koopman spectrum of linear non-autonomous systems from local propagators estimated by DMD on moving stencils; for continuously varying operators this leaves an error proportional to the rate of change of the eigenvalues, which they remove by a change of observables, and their analysis concerns operators whose matrices commute at different times, the periodic case being referred to Mezi\'c and Surana~\cite{Mezic-2016}.
FD addresses the complementary case of an operator that is periodic but otherwise arbitrary, including the non-commuting variation that produces parametric resonance: snapshots one period apart are exact iterates of the monodromy operator, so one regression over all periods suffices, at the price of no time-resolved spectrum within the period and no treatment of aperiodic time dependence.

A sliding-window variant of DMD, applied by the author to nonlinear oscillators before the development of FD \cite[Appendix~D]{Quinlan-2015}, showed the behavior that their analysis explains: accurate estimates where the eigenvalues of the state-dependent system matrix were momentarily stationary and errors elsewhere, in the real part where the frequency varied and in the imaginary part where the growth rate varied.
The same structure appears in the error formula of Ma\'ce\v{s}i\'c et al.~\cite[Theorem~3.1]{Macesic-2018}, which is proved for linear systems in a commuting canonical form.
That experience motivated exploiting periodicity instead of local constancy.

\emph{Slowly varying operators and separated time scales.}
Online and windowed DMD \cite{Zhang-2019} and multiresolution DMD \cite{Kutz-2016-mrDMD} track operators that change slowly or separate time scales; neither addresses an operator that varies periodically on the time scale of the dynamics, where the variation is not a perturbation of a frozen operator but the source of the instability.

\emph{Delay embedding.}
Time-delay coordinates in DMD are also analyzed by Brunton et al.~\cite{Brunton-2017}, Kamb et al.~\cite{Kamb-2020}, and Pan and Duraisamy~\cite{Pan-2020}, and Tu et al.~\cite{Tu-2014} noted that time-shifted copies, as in the eigensystem realization algorithm \cite{Juang-1985}, remove the rank deficiency that defeats DMD for a standing wave; \cref{sec:hankel} identifies what such embeddings return for a periodic system, and \v{C}rnjari\'c-\v{Z}ic et al.~\cite{Crnjaric-2020} give a stochastic Hankel-DMD for noisy data.

\emph{Periodic statistics.}
Phase-locked averaging and the triple decomposition \cite{Reynolds-1972} and the cyclostationary form of spectral proper orthogonal decomposition \cite{Heidt-2024} characterize flows with periodic statistics, but they return no multipliers and so do not address stability.

\emph{Identification of periodic systems.}
Subspace identification has been extended to periodic systems by lifting the input--output data over one period and factoring the resulting block-Hankel matrices \cite{Verhaegen-1995,Hench-1995}, and Floquet multipliers and periodic mode shapes of rotors and wind turbines are identified from records of structural models \cite{Bucher-2001,Allen-2011}.
These are the closest relatives of EFD, which is a DMD counterpart of them, working directly on high-dimensional snapshots with a rank truncation and returning spatial modes; the relation to the stride-one embedding and to the Koopman lattice established above does not appear to have been stated there.
When the governing equations are available, Floquet analysis is carried out on the linearized operator itself \cite{Barkley-1996}.

\emph{Stroboscopic sampling with a known drive.}
Goldschmidt et al.~\cite{Goldschmidt-2021} relate DMD on stroboscopically sampled data of periodically driven quantum systems to Floquet theory, with the drive an input known to the algorithm, the closest published link between DMD and Floquet theory that we know of.
In the systems considered here the periodicity is not an imposed input but arises from the coupling of the parts of the system, and the period must be estimated from the data.

%% file: sections/05_verification.tex
\section{Verification}
\label{sec:verification}

We verify the results of \cref{sec:methods,sec:relations} on systems whose multipliers are known, examine what the methods return when their hypotheses fail, and then consider the choice of the analysis period.
Trajectories are computed with an explicit Runge--Kutta method of order eight at relative and absolute tolerances of $10^{-12}$ and $10^{-14}$, and reference multipliers are obtained by integrating the fundamental matrix over one period at the same tolerances.
The decompositions use the singular value decomposition of the data matrix itself, not the inner-product form of \cref{sec:algorithm}.

\subsection{Linear Mathieu equation}
\label{sec:verif-mathieu}

\Cref{tab:mathieu} lists three parameter sets of the Mathieu equation \cref{eq:mathieu} near its first instability tongue, two inside it and one just outside \cite{Quinlan-2015}.
In Cases 1 and 2 the multipliers are real and negative, so the response is locked at half the modulation frequency and grows (\cref{fig:strobe} shows Case 1); in Case 3 they lie on the unit circle.

\begin{table}[tbhp]
\caption{Floquet multipliers and exponents of the Mathieu equation \cref{eq:mathieu}, $T=2\pi$, from the monodromy matrix (upper part), and the largest relative error of the multipliers estimated from one trajectory started at $(u,\dot u)=(1,0)$ (lower part).
``FD, nearest $p$'': $\dt=10^{-3}$, $50$ time units, $p=6283$ although $T/\dt = 6283.19$.
``FD'', ``EFD'': commensurate sampling with $p=64$, eight periods.
``EFD, scalar'': the same with the single observable $u$.
``DMD'': standard DMD of the full state at the same $\dt$, eigenvalues raised to the power $p$.}
\label{tab:mathieu}
\centering
\small
\begin{tabular}{@{}lccc@{}}
\toprule
 & Case 1 & Case 2 & Case 3 \\ \midrule
$\kappa$ & $0.25$ & $0.25$ & $0.26$ \\
$\varepsilon$ & $0.1$ & $0.01$ & $0.01$ \\
$\mu_1$ & $-1.36711536$ & $-1.03191310$ & $-0.99854406+0.05394220\,\ii$ \\
$\mu_2$ & $-0.73146717$ & $-0.96907385$ & $-0.99854406-0.05394220\,\ii$ \\
$\rho_{1,2}$ & $\pm 0.0497682+0.5\,\ii$ & $\pm 0.0049998+0.5\,\ii$ & $\pm 0.4914107\,\ii$ \\
Stability & unstable & unstable & stable \\ \midrule
\multicolumn{4}{l}{\textit{Largest relative error of the estimated multipliers}} \\
FD, nearest $p$ & $8.1\times10^{-4}$ & $1.2\times10^{-4}$ & $1.6\times10^{-4}$ \\
FD & $6.4\times10^{-13}$ & $1.5\times10^{-13}$ & $7.4\times10^{-13}$ \\
EFD & $1.9\times10^{-13}$ & $2.6\times10^{-13}$ & $7.7\times10^{-14}$ \\
EFD, scalar & $2.9\times10^{-13}$ & $3.0\times10^{-13}$ & $7.0\times10^{-14}$ \\
DMD & $0.93$ & $0.050$ & $0.035$ \\ \bottomrule
\end{tabular}
\end{table}

With commensurate sampling, FD and EFD recover both multipliers to the accuracy of the integration, as \cref{sec:fd,sec:efd} assert, and the stacked EFD matrix has numerical rank two: its third singular value is below $10^{-12}$ of the first.
With the single observable $u$ the requirement of FD fails, since one number cannot determine a two-dimensional state, and FD returns one real eigenvalue that is not a multiplier ($-1.333$, $-1.004$, and $-0.988$ in the three cases).
The system is observable over a period from $u$ alone, however, and EFD remains exact.
Standard DMD of the full state fails for the reason given in \cref{sec:standard-dmd}: the Floquet modes contain several harmonics, no constant $2\times2$ matrix maps one sample to the next, and the eigenvalues of the least-squares fit, raised to the power $p$, miss the multipliers by $3.5\,\%$ to $93\,\%$.

The row ``FD, nearest $p$'' rounds the period to the nearest sample, as must be done when the sampling interval is fixed independently of the period \cite{Quinlan-2015}; its errors, of order $10^{-4}$ and smaller, are those of the small period error (\cref{sec:strides}), and the estimates remain good.

\begin{figure}
\centering
\includegraphics[width=0.32\textwidth]{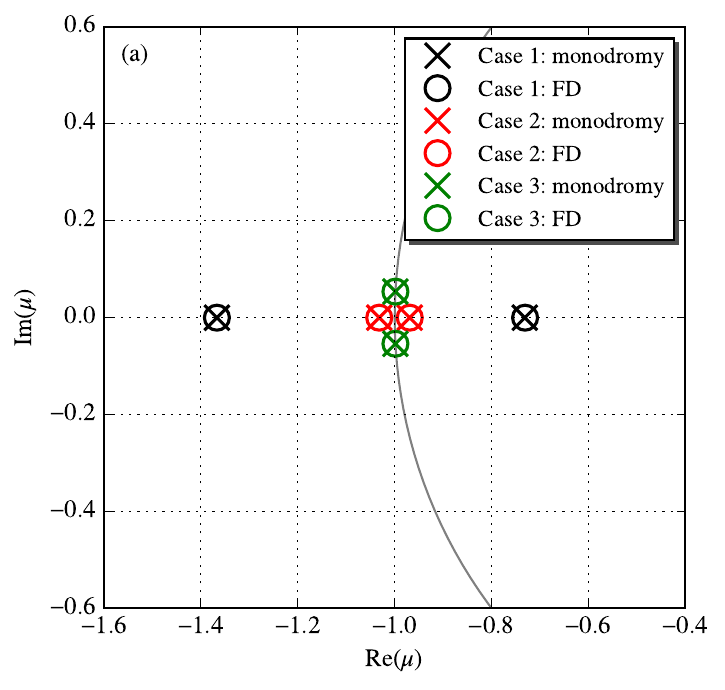}\hfill
\includegraphics[width=0.32\textwidth]{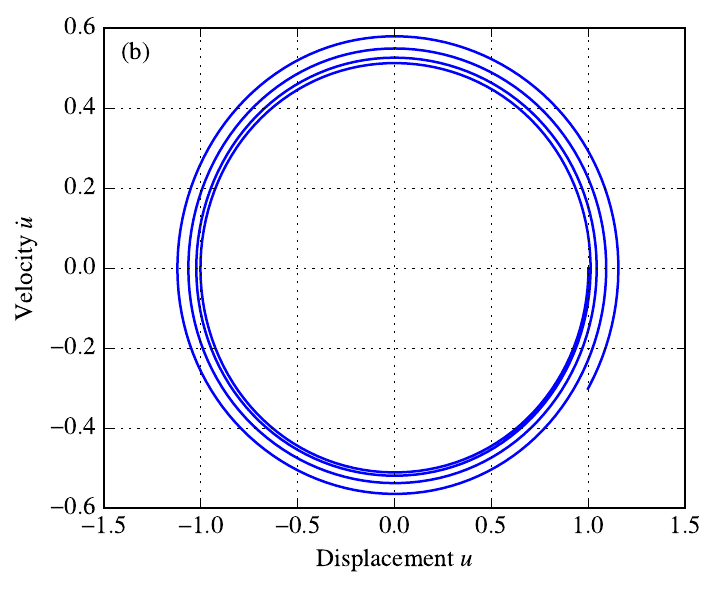}\hfill
\includegraphics[width=0.32\textwidth]{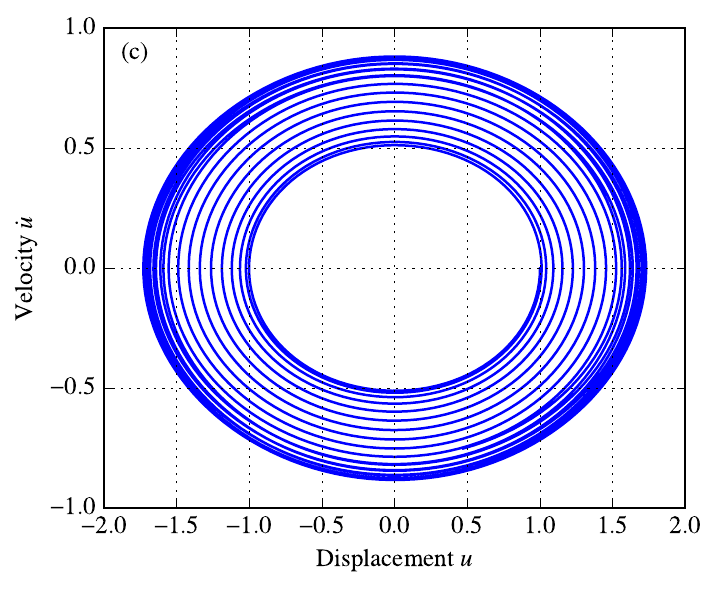}
\caption{Mathieu equation.
(a) Multipliers from the monodromy matrix (crosses) and from FD (circles) for Cases 1--3, with part of the unit circle; FD has the setting of the row ``FD, nearest $p$'' of \cref{tab:mathieu}.
(b), (c) Phase portrait of Case 3 over $50$ and $250$ time units.}
\label{fig:mathieu}
\end{figure}

Case 3 illustrates why multipliers are worth estimating.
The solution is bounded, but it is a slow beat between the frequencies $0.491$ and $0.509$, and over the first fifty time units its amplitude grows (\cref{fig:mathieu}).
The phase portrait must be followed for more than twenty-five periods before the growth is seen to turn, and the power spectra of Cases 2 and 3 over that interval are nearly indistinguishable \cite{Quinlan-2015}.
FD applied to the first fifty time units, fewer than eight periods, returns $\abs{\muh} = 0.99998$.
In the stride-one picture the beat consists of two teeth whose frequencies differ by $0.017$, which a spectrum could separate only from a record several hundred time units long.
At the period they are the complex-conjugate multipliers $\mu_{1,2}$, which a few periods of full-state data determine.

\subsection{The comb of roots}
\label{sec:verif-comb}

\subsubsection{A system with two harmonics}
\label{sec:verif-rotating}

Let $\vx(t) = \mR(\omega t)\,\ee^{\mB t}\vz_0$ with $\mR(\theta)$ the rotation of the plane by the angle $\theta$, $\omega = 2\pi$, $\vz_0 = (1,0.3)^{\trans}$, and
\begin{equation}
  \mB = \begin{bmatrix} -0.05 & 1.7 \\ -2.9 & 0.11 \end{bmatrix}, \qquad \spec(\mB) = \{\,0.03\pm2.2189\,\ii\,\} .
  \label{eq:two-harmonic}
\end{equation}
This is a linear periodic system in Floquet form with $\mP(t) = \mR(\omega t)$ and $T=1$, and $\mu_{1,2} = \ee^{0.03\pm2.2189\ii} = -0.6221\pm0.8215\,\ii$; since $\mR(\omega t)$ contains only the harmonics $\ee^{\pm\ii\omega t}$, each Floquet mode has exactly two harmonics, and \cref{sec:hankel} predicts four teeth, $\ee^{(0.03\pm2.2189\ii\pm\ii\omega)\dt}$.

With $p=10$ and twelve periods of full-state data, Hankel-DMD with $d=2$, $3$, and $10$ has numerical rank four and returns the four teeth, and FD and EFD return $\mu_{1,2}$, all with errors below $5\times10^{-15}$; the stacked EFD matrix has rank two, and the identity \cref{eq:collapse-identity} holds to $3\times10^{-14}$.
Standard DMD of rank two misses the multipliers by $6\,\%$.
The harmonics of each aligned EFD mode are non-zero at $j=\pm1$ only and so identify the lattice frequencies of the mode, and with the strides $s=9$ and $11$ and the window $d=s$ the four eigenvalues equal $\beta_{l,j}^{\,s}$ to $3\times10^{-14}$, as \cref{sec:strides} describes.

\subsubsection{The Mathieu equation}

For the Mathieu equation the comb is infinite, and the result of \cref{sec:hankel} is approached through truncation.
\Cref{tab:comb} shows Case 3 with $p=16$ and forty periods of full-state data, for which the harmonic amplitudes $\abs{a_l}\norm{\vpsih_{l,j}}$ fall off geometrically, by a factor of roughly $250$ for each successive pair of harmonics.
Every eigenvalue returned lies on the predicted comb, with an error that decreases as the window and the rank grow, until it reaches the accuracy of the integration, and that is largest for the weakest teeth retained: the comb of roots is confirmed, tooth by tooth, as far as the arithmetic allows.
FD and EFD at rank two return the multipliers to $10^{-13}$ or better from the same data, an accuracy that Hankel-DMD reaches for its two leading teeth only with the two largest windows and ranks of \cref{tab:comb}.

\begin{table}[tbhp]
\caption{Hankel-DMD of the Mathieu equation, Case 3, $p=16$: error $\abs{\betah-\beta_{l,j}}$ of the eigenvalue nearest each predicted tooth, for several windows $d$ and ranks $r$.
Teeth are listed by harmonic amplitude; each row stands for a complex-conjugate pair.
A dash marks a tooth beyond the truncation rank.}
\label{tab:comb}
\centering
\small
\begin{tabular}{@{}ccccc@{}}
\toprule
 & \multicolumn{4}{c}{Absolute error of the eigenvalue, $\abs{\betah-\beta_{l,j}}$} \\ \cmidrule(l){2-5}
Amplitude & $d=32$, $r=8$ & $d=32$, $r=12$ & $d=48$, $r=16$ & $d=64$, $r=20$ \\ \midrule
$7.7\times10^{-1}$ & $4\times10^{-9}$ & $3\times10^{-12}$ & $7\times10^{-15}$ & $8\times10^{-15}$ \\
$2.1\times10^{-1}$ & $6\times10^{-8}$ & $2\times10^{-11}$ & $6\times10^{-15}$ & $3\times10^{-14}$ \\
$3.1\times10^{-3}$ & $5\times10^{-5}$ & $2\times10^{-9}$ & $7\times10^{-13}$ & $1\times10^{-12}$ \\
$8.4\times10^{-4}$ & $3\times10^{-4}$ & $2\times10^{-8}$ & $5\times10^{-12}$ & $4\times10^{-12}$ \\
$3.8\times10^{-6}$ & --- & $1\times10^{-5}$ & $6\times10^{-10}$ & $3\times10^{-10}$ \\
$1.1\times10^{-6}$ & --- & $8\times10^{-5}$ & $3\times10^{-9}$ & $1\times10^{-9}$ \\
$2.1\times10^{-9}$ & --- & --- & $2\times10^{-6}$ & $7\times10^{-7}$ \\
$6.0\times10^{-10}$ & --- & --- & $9\times10^{-6}$ & $3\times10^{-6}$ \\ \bottomrule
\end{tabular}
\end{table}

\subsection{A limit cycle}
\label{sec:verif-vdp}

For a nonlinear test with known multipliers we take the van der Pol oscillator $\ddot u - b(1-u^2)\dot u + u = 0$ with $b = 0.1$.
Its limit cycle has period $T = 6.287111$, and its multipliers are $\mu_1 = 1$ and $\mu_2 = \exp\int_0^Tb(1-u^2)\,\dd t = 0.533069$, the second by Liouville's formula.
Trajectories are started at the point of the cycle where the displacement is largest, with both coordinates scaled by $1+\epsilon$, sampled with $p=20$, and followed for ten periods.

\Cref{tab:vdp} shows the two FD eigenvalues.
The first is the eigenvalue one of \cref{sec:base-state}, whose mode is the limit cycle itself; the second converges to $\mu_2$ in proportion to $\epsilon$, as expected of a method that is exact for the linearization.
The error at finite amplitude comes from the nonlinearity of the period map near the cycle, whose Koopman eigenvalues include the powers $\mu_2^2,\mu_2^3,\dots$ of the transverse multiplier \cite{Mezic-2016}; a fit of rank two must absorb them into $\muh_2$.
EFD has room for them in its stacked snapshots: at rank four with $\epsilon = 0.02$ it returns $0.533066$, $0.2835$, and $0.139$ against $\mu_2 = 0.533069$, $\mu_2^2 = 0.2842$, and $\mu_2^3 = 0.1515$, and at $\epsilon = 0.2$ it returns $0.5316$ for $\mu_2$ where rank two gives $0.431$.
The advantage is room in the rank, not a different treatment of the nonlinearity; with field data, which have far more observables than retained modes, FD has the same room.

\begin{table}[tbhp]
\caption{Van der Pol limit cycle, $b=0.1$: FD eigenvalues from a trajectory started a relative distance $\epsilon$ off the cycle.
The exact multipliers are $1$ and $0.533069$.}
\label{tab:vdp}
\centering
\small
\begin{tabular}{@{}cccc@{}}
\toprule
$\epsilon$ & $\muh_1$ & $\muh_2$ & Relative error of $\muh_2$ \\ \midrule
$2\times10^{-1}$ & $0.998036$ & $0.403247$ & $2.4\times10^{-1}$ \\
$2\times10^{-2}$ & $0.999974$ & $0.518955$ & $2.6\times10^{-2}$ \\
$2\times10^{-3}$ & $0.9999997$ & $0.531647$ & $2.7\times10^{-3}$ \\
$2\times10^{-4}$ & $1.0000000$ & $0.532927$ & $2.7\times10^{-4}$ \\ \bottomrule
\end{tabular}
\end{table}

\subsection{Beyond the hypotheses: a conservative nonlinear system}
\label{sec:verif-nonlinear}

The results above concern linear periodic systems and small perturbations of periodic orbits; measured data seldom satisfy either description exactly, and it is as important to know what the methods return when they do not.
Consider the pair of coupled oscillators
\begin{equation}
  \ddot u_1 + (\kappa + \varepsilon u_2)\,u_1 + \varepsilon\alpha\,u_1^3 = 0, \qquad
  \ddot u_2 + u_2 = \tfrac12\,\varepsilon\,u_1^2 ,
  \label{eq:coupled}
\end{equation}
with $\kappa = 0.25$ and $\varepsilon = \alpha = 0.1$, a variant of a system proposed by Holmes and Rand~\cite{Holmes-1981} as an autonomous analog of a nonlinear Mathieu equation: the second oscillator, near unit frequency, modulates the stiffness of the first, which responds near half that frequency and feeds back quadratically \cite{Quinlan-2015}.
The system is autonomous, so nothing in it is periodic a priori, and it is nonlinear; from $(u_1,u_2,\dot u_1,\dot u_2) = (1,1,0,0)$ the solution is a bounded, sustained, modulated oscillation.

\emph{Moduli are not decay rates here.}
With the analysis period $T = 6.261$ estimated from the autocorrelation of the state (\cref{sec:period}), FD of the full state over $150$ time units returns two complex pairs, of moduli $0.939$ and $0.943$ \cite{Quinlan-2015}.
The vector field of \cref{eq:coupled} is divergence-free, so its flow preserves volume and no solution is asymptotically stable, and the solution shows no decay over $600$ time units; yet the moduli lie between $0.90$ and $0.99$ for every window of $150$ time units within that record.
The contraction is a property of the fit, not of the system, and we attribute it to model mismatch: a linear model of rank four cannot represent a trajectory whose frequencies depend on amplitude, the part of each snapshot that the model fails to predict enters the regression as noise does, and least-squares estimates of DMD eigenvalues from noisy data shrink toward the origin \cite{Dawson-2016,Hemati-2017}.
A modulus below one obtained from a signal that does not decay measures the loss of coherence from one period to the next and must not be read as a decay rate (\cref{sec:experiment}).

\emph{Partial observation.}
From the two displacements $(u_1,u_2)$ alone, FD at the same period returns two real eigenvalues, since two observables cannot represent two complex pairs, whereas EFD of the same data returns both pairs, $-0.862\pm0.399\,\ii$ and $0.912\pm0.262\,\ii$, within about $3\,\%$ of the full-state values.
The velocities need not be measured, because the displacements, sampled through one period, determine them: the content of \cref{sec:efd}, carried over to a system that satisfies its hypotheses only approximately.

\subsection{Choosing the analysis period}
\label{sec:period}

FD and EFD require the period, or rather the integer $p$.
When the periodicity is imposed, $T$ is known; otherwise it is estimated from the data before the decomposition, from the dominant narrow peak in the spectrum of a long record, checked against the autocorrelation and against a weakly damped mode of a standard DMD of the same data \cite{Quinlan-2015}, and $p = \mathrm{round}(T/\dt)$.
How much an error in the period matters depends on the regime of \cref{sec:strides}.
Rounding to the nearest sample costs little (\cref{tab:mathieu}).
A poor estimate analyzed at minimal rank costs much: for the displacement data of \cref{eq:coupled}, FD at minimal rank with a period $8\,\%$ too short changed the modulus from $0.636$ to $0.763$ \cite{Quinlan-2015}.
With a window and spare rank the estimates tolerate errors of several percent.
For Case 1 of the Mathieu equation, sampled at $\dt = T/200$ and analyzed with strides up to $8\,\%$ away from the period, $\mathrm{DMD}(s,s)$ at rank $6$ returned the growth rate of the unstable multiplier within $2.1\,\%$ at every stride, against $6.9\,\%$ for FD at rank $2$, and in the experiment of \cref{sec:experiment} EFD with analysis periods of $9$, $10$, and $11$ samples gave frequencies of the leading mode within $1\,\%$ of one another.
The robustness is therefore to the discretization of a good estimate, not to a poor estimate analyzed at minimal rank.
When two periodicities are present, any common period is admissible in principle, since a $T$-periodic system is also $2T$-periodic, but the longer period leaves fewer periods in the record and folds the frequencies into a narrower band \cref{eq:unwrap}.

%% file: sections/06_symmetry.tex
\section{Periodicity induced by symmetry and observation}
\label{sec:symmetry}

A time-periodic linear operator need not signal a parametric excitation.
It also arises when an autonomous flow with a rotational symmetry has a rotating wave as its base state, the situation of a precessing vortex core or a helical vortex in a swirling jet \cite{Syred-2006,Oberleithner-2011}.
This section shows that the two cases lead to the same data model, identifies the multipliers in the second case, and explains why the periodicity can be removed from full-state data but not from a measurement in a fixed plane.

\subsection{A rotating wave in two frames}
\label{sec:rotating}

Let $\dot{\vu} = \vf(\vu)$ be an autonomous system on $\RR^n$ that is equivariant under rotation about an axis: $\vf(\mR_\theta\vu) = \mR_\theta\vf(\vu)$, where $\mR_\theta = \ee^{\theta\mJ}$ rotates the state by the angle $\theta$ and $\mR_{2\pi} = \mI$ \cite{Chossat-2000}.
Rotating a state and then advancing it in time gives the same result as advancing it and then rotating.
A \emph{rotating wave} is a solution of the form $\vu(t) = \mR_{\Omega t}\vu_*$ with $\Omega\ne0$; it is a steady state in a frame that rotates at the rate $\Omega$, and $\vu_*$ satisfies $\vf(\vu_*) = \Omega\mJ\vu_*$.
Let $q\ge1$ be the order of the discrete symmetry of the wave, that is, the largest integer with $\mR_{2\pi/q}\vu_* = \vu_*$.

\emph{In the co-rotating frame} write $\vu = \mR_{\Omega t}(\vu_*+\vw)$.
Substituting into $\dot{\vu} = \vf(\vu)$ and using the equivariance,
\begin{equation}
  \dot{\vw} = \vf(\vu_*+\vw) - \Omega\mJ(\vu_*+\vw) = \mL\vw + O(\abs{\vw}^2), \qquad \mL = \mathrm{D}\vf(\vu_*) - \Omega\mJ ,
  \label{eq:corotating}
\end{equation}
where $\mathrm{D}\vf$ is the Jacobian matrix of $\vf$.
The linearized equation is autonomous, and the eigenvalues of $\mL$ decide the stability of the wave.

\emph{In the laboratory frame} the perturbation is $\vu' = \vu - \mR_{\Omega t}\vu_* = \mR_{\Omega t}\vw$.
Differentiating the equivariance condition gives $\mathrm{D}\vf(\mR_\theta\vu)\mR_\theta = \mR_\theta\mathrm{D}\vf(\vu)$, and with it, to first order,
\begin{equation}
  \dot{\vu}' = \mA(t)\,\vu', \qquad \mA(t) = \mathrm{D}\vf(\mR_{\Omega t}\vu_*) = \mR_{\Omega t}\,\mathrm{D}\vf(\vu_*)\,\mR_{-\Omega t}, \qquad \mPhi(t,0) = \mR_{\Omega t}\,\ee^{\mL t} .
  \label{eq:lab-frame}
\end{equation}
Because $\mR_{2\pi/q}\vu_* = \vu_*$, the operator $\mA(t)$ is periodic with period $T = 2\pi/(q\Omega)$: the laboratory sees the same base state again after the wave has turned through one $q$-th of a revolution.

\emph{The multipliers} follow from the monodromy matrix $\mM = \mPhi(T,0) = \mR_{2\pi/q}\,\ee^{\mL T}$.
The symmetry of $\vu_*$ gives $\mathrm{D}\vf(\vu_*)\mR_{2\pi/q} = \mR_{2\pi/q}\mathrm{D}\vf(\vu_*)$, so $\mR_{2\pi/q}$ commutes with $\mL$ and with $\ee^{\mL T}$.
Commuting matrices can be brought to triangular form together, so the eigenvalues of their product are products of paired eigenvalues, and those of $\mR_{2\pi/q}$ are $q$-th roots of unity.
The Floquet multipliers are therefore
\begin{equation}
  \mu = \ee^{\lambda T}\,\ee^{2\pi\ii\,\kappa/q} , \qquad \lambda\in\spec(\mL),
  \label{eq:rotating-multipliers}
\end{equation}
where $\ee^{2\pi\ii\kappa/q}$, with $\kappa$ an integer modulo $q$, is the eigenvalue of $\mR_{2\pi/q}$ on the corresponding common eigenvector.
In particular $\mu = \ee^{\lambda T}$ when $q=1$.

\emph{The data.}
If the perturbation is observed through a constant matrix $\mC$ in the laboratory frame, $\snap_k = \mC\vu'_k$, the data satisfy the model \cref{eq:sampled} with $\mC_k = \mC$ and $\mPhi_k = \mPhi(t_{k+1},t_k)$, and the results of \cref{sec:fd,sec:efd,sec:hankel} apply with the multipliers \cref{eq:rotating-multipliers}.

\subsection{What a fixed measurement plane can and cannot do}
\label{sec:fixed-plane}

The two frames describe one system, an instance of the equivalence of \cref{sec:sampled}: in the co-rotating frame the dynamics are constant and a fixed laboratory sensor is the periodic observation $\mC\mR_{\Omega t}$, and in the laboratory frame the sensor is constant and the dynamics are periodic, the periodic change of coordinates of Floquet's theorem being the rotation itself.

If the full state is measured, the rotation can be undone by a template-fitting or phase-reduction step that maps each snapshot to the co-rotating frame \cite{Rowley-2000,Marensi-2023}, and standard DMD of the reduced data estimates $\spec(\mL)$ directly.
A measurement in a fixed plane or at fixed points, such as the planar velocity data in which the helical vortices near $1$\,kHz in the combustor of \cref{sec:experiment} were identified \cite{Quinlan-2017}, does not contain the information needed to rotate it: the observation does not create the periodic operator, but it removes the option of eliminating it by a change of frame.
A well-behaved autonomous flow seen through a fixed plane then presents data that no constant operator fits and that Floquet theory describes exactly, with multipliers that are, by \cref{eq:rotating-multipliers}, the exponentials of the co-rotating eigenvalues over one period; the system of \cref{sec:verif-rotating} is the smallest instance, for which FD fails, EFD is exact, and Hankel-DMD returns the two-tooth combs $\ee^{(\lambda\pm\ii\Omega)\dt}$.

Two remarks bear on the interpretation of data.
First, $\mJ\vu_*$ is an eigenvector of $\mL$ with eigenvalue zero, the infinitesimal rotation of the wave, so one multiplier of a rotating wave is always one, in addition to the eigenvalue one of the base state (\cref{sec:base-state}).
Second, by \cref{sec:sampled}, data from a fixed plane cannot distinguish a rotating structure from a parametric modulation at the same period; the method returns the same kind of answer in both cases, which is its strength, and deciding between them is a matter of physics.

%% file: sections/07_experiment.tex
\section{Application to measurements in a swirl-stabilized combustor}
\label{sec:experiment}

We demonstrate FD and EFD on measured data whose governing operator is unknown, and compare them with DMD and Hankel-DMD on the same data.
The physical interpretation of the modes found and the mechanism of the instability are the subject of a companion paper \cite{Quinlan-2026-PCI}, of which a preliminary account appeared in Quinlan~\cite{Quinlan-2022}, and which suggests that the instability is a parametric, acoustically coupled combustion instability; here the data serve to show what the methods return, how EFD differs from FD on noisy, partial measurements, and how sensitive the results are to the choices the analyst must make.

\subsection{Data}
\label{sec:exp-data}

The facility is a laboratory combustor with a single swirl injector that burns hydrogen, doped with methane, in air, inside a duct whose transverse length can be adjusted \cite{Quinlan-2017,Quinlan-2015}; the companion paper \cite{Quinlan-2026-PCI} describes it and the measurements in detail.
Two configurations are considered: in the \emph{unstable} configuration the dynamic pressure spectrum has a narrow peak at $169.7$\,Hz, and in the \emph{stable} configuration it does not.
The measurements are simultaneous and time-resolved at $10$\,kHz: planar particle image velocimetry (two velocity components in a plane through the injector axis), planar laser-induced fluorescence of OH, line-of-sight chemiluminescence of CH$^*$, and three dynamic pressure signals.
In the snapshot, the dynamic pressure signals are band-pass filtered between $80$ and $500$\,Hz without phase shift and accompanied by their time derivatives; the band keeps the two peaks studied below and excludes the band near $1$\,kHz that dominates the spectra of the sensors near the injector.
The band-pass is a choice of observable: it moves no eigenvalue of a linear system, but it is a prior selection of frequencies, and \cref{sec:exp-sensitivity} shows how the methods depend on its cutoff.

Each snapshot $\snap_k$ is the concatenation of all of these diagnostics at one instant, as in Richecoeur et al.~\cite{Richecoeur-2012}, so that a decomposition of the composite snapshots yields modes whose velocity, scalar, and dynamic pressure parts share one eigenvalue.
Each kind of data is multiplied by one factor, chosen so that the norm of its time-averaged absolute value equals a prescribed number (\cref{sec:algorithm}); the baseline norms are $1/3$, $1$, $1$, and $3$ for the velocity, OH, CH$^*$, and dynamic pressure data, the dynamic pressure being weighted most because it alone is free of aliasing and the instability is acoustic, and the velocity least because it is the noisiest.
The baseline record has $N = 8504$ snapshots, and all decompositions use the rank $r = 25$ unless stated otherwise.
The planar and line-of-sight measurements observe a three-dimensional swirling reacting flow only in part.

\subsection{Analysis period}
\label{sec:exp-period}

The analysis period is chosen as the period of the narrowband activity of the injector shear layer near $1$\,kHz \cite{Quinlan-2017,Quinlan-2026-PCI}, which is not a sharp period.
Applied to the dynamic pressure, the three estimators of \cref{sec:period} give, for the unstable configuration, a spectral peak near $1000$\,Hz, an autocorrelation maximum at a lag of $1.01$\,ms ($989.9$\,Hz), and, with the dynamic pressure signals not high-pass filtered, a DMD mode at $1017$\,Hz \cite{Quinlan-2015}.
These rest on the sensors near the injector, whose mounting tubes resonate in the same band, so they locate the band rather than measure a period.
The band is that of a hydrodynamic mode of the injector shear layer, a pair of helical vortices found by DMD of the velocity and scalar data at $984$\,Hz \cite{Quinlan-2017}; it is present without combustion and is narrowband rather than periodic \cite{Quinlan-2026-PCI}.
We use $p=10$ samples, $T=1$\,ms, for both configurations, and \cref{sec:exp-sensitivity} shows that the results depend little on this choice within $\pm10\,\%$.
The record then contains $850$ whole periods, and the EFD data matrix has $K = 850$ columns, one for each period.

\subsection{Four decompositions of one record}
\label{sec:exp-results}

\Cref{tab:exp-methods} lists what the four methods of \cref{tab:methods} return near the two frequencies at which the dynamic pressure spectrum of the unstable configuration has narrow peaks, $170$ and $258$\,Hz.
To make the methods comparable, the eigenvalues of the two methods with stride one are raised to the power $p$, so that every entry is the change over one analysis period.

\begin{table}[tbhp]
\caption{The eigenvalue of largest modulus within $8$\,Hz of each of the two peaks of the dynamic pressure, unstable configuration, baseline scaling, $p = 10$, rank $25$: frequency and modulus per analysis period.
For DMD and Hankel-DMD the modulus is $\abs{\betah}^{\,p}$, the change over one period.
$d$ is the window and $s$ the stride; a dash means that no eigenvalue lies within $8$\,Hz of the peak.}
\label{tab:exp-methods}
\centering
\small
\begin{tabular}{@{}lcccc@{}}
\toprule
Method & $d$ & $s$ & First pair & Second pair \\ \midrule
DMD & $1$ & $1$ & --- & --- \\
FD, every section & $1$ & $10$ & --- & --- \\
Hankel-DMD & $10$ & $1$ & $167.74$\,Hz, $0.9209$ & $263.70$\,Hz, $0.9726$ \\
EFD & $10$ & $10$ & $171.32$\,Hz, $0.9530$ & $262.48$\,Hz, $0.9524$ \\ \bottomrule
\end{tabular}
\end{table}

Standard DMD has no eigenvalue within $8$\,Hz of either frequency, in agreement with Quinlan~\cite{Quinlan-2015}.

FD uses one snapshot per period, so the record contains ten disjoint FD data sets, one for each Poincar\'e section $c = 1,\dots,10$, and since a mode that is weakly observed at one phase could be missed there and found at another, all ten were analyzed.
None has an eigenvalue within $8$\,Hz of either pair, so the failure is not a matter of observability at a particular section.
Raising the rank of FD recovers the first pair but not the second: at ranks $50$ to $200$, two to eight times the rank at which EFD finds it, all ten sections return it, with moduli of $0.85$ to $0.91$.

\begin{figure}
\centering
\includegraphics[width=0.32\textwidth]{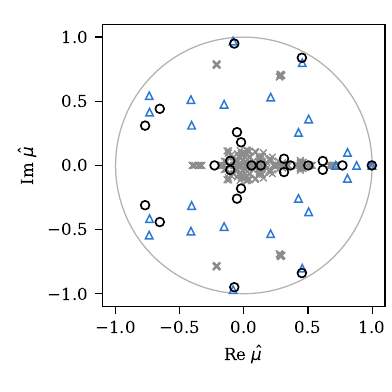}\hfill
\includegraphics[width=0.32\textwidth]{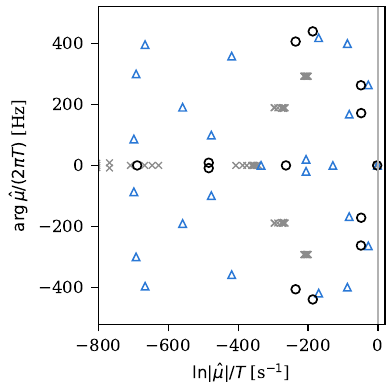}\hfill
\includegraphics[width=0.32\textwidth]{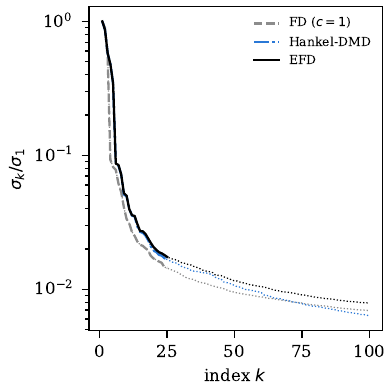}
\caption{Unstable configuration, $T=1$\,ms, rank $25$, baseline scaling.
Left: multipliers of FD on all ten Poincar\'e sections (gray crosses) and of EFD (circles), and the eigenvalues of Hankel-DMD raised to the power $p$ (triangles), with the unit circle.
Center: the same as exponents $\ln(\muh)/T$ on the principal branch.
Right: normalized singular values of the FD (section $c=1$), Hankel-DMD, and EFD data matrices; the heavy portions mark the $25$ retained.}
\label{fig:exp-eigs}
\end{figure}

EFD and Hankel-DMD both find the two pairs, close to the unit circle and at frequencies within $4$\,Hz of each other.
EFD places them at $171.32$\,Hz with $\abs{\muh} = 0.9530$ and at $262.48$\,Hz with $\abs{\muh} = 0.9524$.
Hankel-DMD, on the same block-Hankel matrix with stride one, returns $167.74$\,Hz with $\abs{\betah}^{\,p} = 0.9209$ and $263.70$\,Hz with $\abs{\betah}^{\,p} = 0.9726$.
\Cref{fig:exp-eigs} shows the full spectra.
For the two pairs, the Hankel-DMD eigenvalues raised to the power $p$ fall near the EFD multipliers, as \cref{sec:collapse} predicts for a linear periodic system, while the weaker eigenvalues of the two methods do not correspond one to one, as is to be expected of data far from that setting.
The frequencies lie within $2$\,Hz of the first peak of the measured dynamic pressure spectrum and within $6$\,Hz of the second.

The modes agree as well as the eigenvalues.
By \cref{sec:collapse} the average of the phase-aligned EFD mode over the ten phases is the leading block of the Hankel-DMD mode, and the two agree to $0.94$ or better in every kind of data for both pairs, measured by $\abs{\langle\vb{a},\vb{b}\rangle}/(\norm{\vb{a}}\,\norm{\vb{b}})$, whereas a single, unaveraged block of the EFD mode agrees in velocity only to $0.85$ and $0.63$.
In both EFD modes the harmonics other than zero are weak, the largest having $0.6\,\%$ and $1.1\,\%$ of the norm of harmonic zero, which is why Hankel-DMD shows one eigenvalue for each pair rather than a comb (\cref{sec:hankel}).
On these data EFD and Hankel-DMD are two packagings of the same information, and EFD returns it from a data matrix with one tenth as many columns, with the multipliers directly and the phase dependence of the modes resolved.

In the stable configuration, whose dynamic pressure spectrum has no peak at $170$\,Hz, EFD returns no pair near the unit circle at that frequency, as in Quinlan~\cite{Quinlan-2015}, and a pair at $266$\,Hz with $\abs{\muh} = 0.905$, close in frequency and in structure to the $262$\,Hz pair above.

\subsection{Phase-resolved and averaged modes}
\label{sec:exp-modes}

\begin{figure}
\centering
\includegraphics[width=0.4\textwidth]{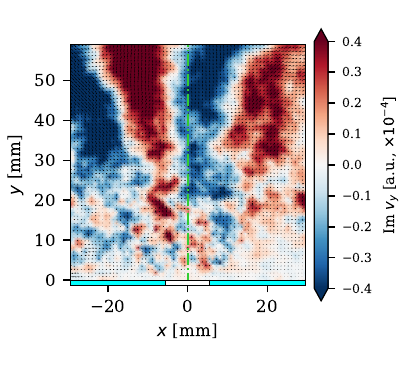}\qquad
\includegraphics[width=0.4\textwidth]{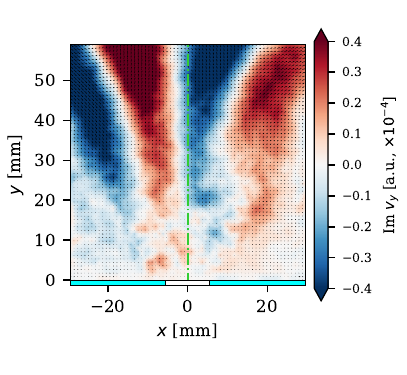}
\caption{Imaginary part of the axial-velocity component of the EFD mode at $171.32$\,Hz, unstable configuration.
Left: the block for the first phase.
Right: the average of the ten phase-aligned blocks.
The coordinate $x$ runs along the duct, positive to the right, as in the companion paper \cite{Quinlan-2026-PCI}; the injector is at the bottom of the field, $y=0$: the strip below the field shows the injector face (cyan) and the $10.9$\,mm nozzle opening (white), and the dash-dot line is the injector centerline.
Arrows show the imaginary part of the in-plane velocity of the mode at every velocity vector; they are small and visible on zooming in.
The mode's arbitrary phase is fixed so that the first dynamic pressure channel of the average is real and negative, and the units are arbitrary.
The values reach $0.8$ in magnitude in both panels, but the color scale is clipped at $\pm0.4$ to bring out the alternating bands.}
\label{fig:exp-mode}
\end{figure}

The EFD mode at $171.32$\,Hz consists of ten blocks, one for each phase of the $1$\,ms analysis period, which is about one sixth of the cycle of the mode, so the blocks resolve the dependence of the mode on the phase of the analysis period, not its oscillation at $170$\,Hz, which the complex mode carries in its real and imaginary parts.
\Cref{fig:exp-mode} shows the axial-velocity part of one block and of the average of the ten aligned blocks (\cref{sec:efd-modes}): the single block shows an organized structure in the upper half of the field and irregular small-scale content near the injector, and the average keeps the former and removes the latter, which is not repeated from phase to phase.

\subsection{Sensitivity}
\label{sec:exp-sensitivity}

\Cref{tab:exp-sensitivity} reports the frequency and modulus of the two leading EFD pairs when the analysis is changed in four ways: the relative scaling of the four kinds of data in the composite snapshot; the analysis period, by one sample in either direction; the truncation rank; and the portion of the record used, where the two halves share no data.
Because a modulus near one conceals large relative changes in the decay rate, the table also lists $\ln\abs{\muh}/T$.

\begin{table}[tbhp]
\caption{Sensitivity of the two leading EFD pairs, unstable configuration.
The baseline (first row) has norms $(1/3,1,1,3)$ for the velocity, OH, CH$^*$, and dynamic pressure data, $p=10$, rank $25$, and all $8504$ snapshots; each other row changes one of these.
Frequencies $f$ in Hz; rates $\ln\abs{\muh}/T$ in s$^{-1}$.
A dash means that no eigenvalue lies within $8$\,Hz of the pair.}
\label{tab:exp-sensitivity}
\centering
\small
\begin{tabular}{@{}lcccccc@{}}
\toprule
 & \multicolumn{3}{c}{First pair} & \multicolumn{3}{c}{Second pair} \\ \cmidrule(lr){2-4}\cmidrule(l){5-7}
Change & $f$ & $\abs{\muh}$ & rate & $f$ & $\abs{\muh}$ & rate \\ \midrule
baseline & $171.32$ & $0.9530$ & $-48.2$ & $262.48$ & $0.9524$ & $-48.8$ \\ \midrule
norms $(1,1,1,1)$ & \multicolumn{3}{c}{---} & \multicolumn{3}{c}{---} \\
norms $(1/10,1,1,1)$ & $175.80$ & $0.8965$ & $-109.2$ & \multicolumn{3}{c}{---} \\
norms $(1,1,1,10)$ & $169.52$ & $0.9749$ & $-25.4$ & $262.91$ & $0.9598$ & $-41.0$ \\
norms $(1/3,1,1,1)$ & $176.05$ & $0.8939$ & $-112.2$ & \multicolumn{3}{c}{---} \\
norms $(1/3,1,1,1/10)$ & \multicolumn{3}{c}{---} & \multicolumn{3}{c}{---} \\
norms $(1/3,1,1,0)$ & \multicolumn{3}{c}{---} & \multicolumn{3}{c}{---} \\ \midrule
$p=9$ ($T=0.9$\,ms) & $172.12$ & $0.9423$ & $-66.1$ & $263.00$ & $0.9403$ & $-68.3$ \\
$p=11$ ($T=1.1$\,ms) & $170.71$ & $0.9598$ & $-37.3$ & $262.28$ & $0.9577$ & $-39.3$ \\ \midrule
rank $10$ & $174.43$ & $0.9062$ & $-98.5$ & \multicolumn{3}{c}{---} \\
rank $15$ & $172.24$ & $0.9419$ & $-59.9$ & $265.72$ & $0.9209$ & $-82.4$ \\
rank $100$ & $169.55$ & $0.9781$ & $-22.1$ & $262.90$ & $0.9638$ & $-36.8$ \\ \midrule
snapshots $1$--$4000$ & $172.54$ & $0.9315$ & $-71.0$ & $260.89$ & $0.9505$ & $-50.7$ \\
snapshots $4001$--$8000$ & $170.92$ & $0.9667$ & $-33.8$ & $264.69$ & $0.9609$ & $-39.9$ \\ \bottomrule
\end{tabular}
\end{table}

The two pairs are found in every case with the dynamic pressure data weighted at least three times as heavily as the scalar data and a rank of at least $15$, including the two disjoint halves of the record, which share neither measurement noise nor turbulent fluctuations, and the frequency of each pair then varies by at most $1.8\,\%$; a change of $\pm10\,\%$ in the analysis period changes that of the first pair by less than $0.5\,\%$, as \cref{sec:strides} predicts for an analysis with a delay window and spare rank.
Hankel-DMD is less consistent: in the same cases it misses the first pair in the second half of the record.
With the dynamic pressure weighted no more heavily than the scalar data the pairs are weakened or lost, and the velocity and scalar data alone return neither.
The two pairs are therefore identified through the dynamic pressure, and their velocity and scalar parts are the parts of those fields that are coherent with it; their physical interpretation is the subject of the companion paper \cite{Quinlan-2026-PCI}.

The band of the dynamic pressure matters too, and it separates the methods.
With the low-pass cutoff of both dynamic pressure signals varied between $300$\,Hz and $8$\,kHz at rank $25$, all four methods find both pairs at $300$\,Hz; DMD and FD lose the first pair above about $400$\,Hz, EFD keeps it up to $664$\,Hz with a modulus that falls steadily, Hankel-DMD finds and misses it in turn between $375$ and $688$\,Hz, and no method finds it at $1$\,kHz or above.
The contrast of \cref{tab:exp-methods} between EFD and the two methods without a window therefore holds over a range of cutoffs, about $420$ to $560$\,Hz, not only at the baseline, and of the two methods with a window EFD is the more consistent.

The window of Hankel-DMD is a choice as well.
It has been the period, $d = p$, with which Hankel-DMD carries the same information as EFD (\cref{sec:collapse}), but an analyst who did not know the period would choose some other window.
With stride one and windows of $1$ to $30$ samples at ranks $25$ and $50$, both pairs are found only with windows of at least about the period, $d\ge9$ at rank $25$, with moduli that rise with the window.
Shorter windows return the first mode $8$ to $26$\,Hz too high and miss the second: the eigenvalue returned is a mixture of the two modes, in which the pattern of the three dynamic pressure signals contains the second mode at $16$ to $24\,\%$ of the fitted amplitude, against at most $10\,\%$ where the first pair is found.
The threshold is a time, not a number of samples: with every second snapshot, at $5$\,kHz, both pairs are found from $d = 6$, and nearly at $d = 5$, about $1$\,ms in both cases, and EFD with $p = 5$ finds both.
The window must therefore span about one cycle of the hydrodynamic mode of the injector shear layer near $1$\,kHz identified earlier \cite{Quinlan-2017}, which sets the analysis period (\cref{sec:exp-period}).
The counting conditions of \cref{sec:hankel} do not decide this, since $d \ge r$ is not needed and $md \ge r$ already holds with $d = 1$: what the window must contain is enough of the state to separate the modes, the condition under which EFD succeeds (\cref{prop:efd}).
Knowing the period thus gives the user of Hankel-DMD the shortest window to use, $d \ge p$; for EFD it is the stride itself.

The moduli are less certain than the frequencies.
Where the first pair is found they lie between $0.894$ and $0.978$, and they move toward the unit circle as the rank is raised, as the dynamic pressure is weighted more heavily, and in the later half of the record, the pattern expected of the bias toward the origin that noise and unmodeled dynamics impose on DMD eigenvalues \cite{Dawson-2016,Hemati-2017} (\cref{sec:algorithm,sec:verif-nonlinear}).
The ordering of the modes by modulus is robust, but the moduli themselves are lower bounds on coherence, not decay rates: the oscillation at $170$\,Hz is sustained throughout the record, whereas a modulus of $0.9530$ per millisecond would, taken literally, have it decay by a factor of $\ee$ in $21$\,ms.
A quantitative stability margin from data of this kind requires a treatment of noise, for which bias-corrected and stochastic variants of DMD \cite{Dawson-2016,Hemati-2017,Crnjaric-2020} offer starting points.

Finally, the analysis period is that of a narrowband motion, not of a strictly periodic flow, and the identification of the two pairs, whose frequencies depend little on the period, holds equally for time-invariant dynamics (\cref{sec:strides}), so it is not evidence that the dynamics are periodic.
The activity near $1$\,kHz is a pair of helical vortices rotating through the fixed plane of the measurements, and the data alone cannot say whether it modulates the dynamics parametrically or is a rotating structure seen in a fixed plane (\cref{sec:sampled,sec:symmetry}); for the acoustic mode, which is fixed in the duct, the two descriptions coincide.
The companion paper \cite{Quinlan-2026-PCI} takes up that question and the mechanism of the instability, and suggests that the instability is a parametric, acoustically coupled combustion instability.

%% file: sections/08_guidance.tex
\section{Recommendations for use}
\label{sec:guidance}

This section condenses the analysis and the examples into a procedure; each item points to where it is established.

\begin{enumerate}
\item[(i)] \emph{Decide whether the method applies.}
FD and EFD estimate the multipliers of dynamics that are periodic, in the operator or in its observation, with a period that is imposed or that of a coherent motion or base state (\cref{sec:sampled}); they return eigenvalues for any analysis period, also for time-invariant data, so the output is not by itself evidence that the dynamics are periodic (\cref{sec:exp-sensitivity}).

\item[(ii)] \emph{Estimate the period first} from a spectrum of a long record, checked against the autocorrelation and a standard DMD, round it to a whole number $p$ of samples, and repeat the analysis with $p\pm1$ (\cref{sec:period}).

\item[(iii)] \emph{Prefer EFD to FD unless the whole state is measured}, as in a simulation: EFD needs only that the measurements made during one period determine the state (\cref{sec:efd}), and in the experiment FD missed the instability on every Poincar\'e section (\cref{sec:exp-results}).

\item[(iv)] \emph{Assemble and scale the snapshot.}
Measurements of different kinds may be concatenated, so that their parts of a mode share one eigenvalue (\cref{sec:exp-data}); state and vary the relative weights and the band of any filter applied before the decomposition (\cref{sec:algorithm,sec:exp-sensitivity}).

\item[(v)] \emph{Choose the rank by its effect}: vary it, and expect the frequencies to hold and the moduli to move (\cref{sec:algorithm,sec:exp-sensitivity}).

\item[(vi)] \emph{Read the output.}
An eigenvalue near one is the periodic base state (\cref{sec:base-state}); the arguments of the multipliers give frequencies modulo $1/T$, with the branch fixed by the dominant harmonic of the EFD mode or by a spectrum of the full-rate data (\cref{sec:leading,sec:efd-modes}); and the average of the aligned blocks of an EFD mode suppresses what is not repeated from phase to phase (\cref{sec:efd-modes,sec:exp-modes}).

\item[(vii)] \emph{Do not read a modulus below one as a decay rate for measured data}: it is biased toward the origin by noise and by dynamics that the model cannot represent, and it is a lower bound on the coherence from one period to the next; the frequencies and the ordering of the modes are the robust outputs (\cref{sec:verif-nonlinear,sec:exp-sensitivity}).

\item[(viii)] \emph{Check the result} on the two halves of the record, with the period changed by one sample, and against Hankel-DMD on the same stacked snapshots, whose eigenvalues raised to the power $p$ should fall on the EFD multipliers and whose leading mode blocks should match the averaged EFD modes (\cref{sec:collapse,sec:exp-results}); with the inner products of the snapshots stored, these checks cost little (\cref{sec:algorithm}).

\item[(ix)] \emph{Know what is not provided}: whether the periodicity is parametric or observational (\cref{sec:sampled,sec:symmetry}), a spectrum resolved within the period beyond the phase dependence of the modes, a treatment of aperiodic time dependence (\cref{sec:other-methods}), and a correction of the moduli for noise.
\end{enumerate}

%% file: sections/09_conclusions.tex
\section{Conclusions}
\label{sec:conclusions}

Schmid's remark that sampling at the period of an oscillatory flow identifies the Floquet matrix \cite{Schmid-2010} has been developed here into a method with stated conditions of validity.
Viewed once per period, a linear time-periodic system is advanced by one constant matrix, the monodromy matrix, whether the periodicity lies in the dynamics or in the observation, and DMD of period-sampled snapshots returns its eigenvalues, the Floquet multipliers, exactly: from a single phase when the measurement at that phase determines the excited state (FD), and from all phases stacked together when the measurements over one period do (EFD).
The second condition is the weaker one, and it is the reason EFD succeeds with scalar and other partial measurements where FD cannot: stacking does not raise the rank of the data, it changes what must be observable.
The EFD mode is the Floquet mode resolved in phase, and its harmonics are the Koopman modes of the lattice of Mezi\'c and Surana~\cite{Mezic-2016}.

On the same block-Hankel matrix, Hankel-DMD with a stride of one sample returns for each multiplier a comb of its $p$-th roots, one for each harmonic present in the observed Floquet mode: the lattice sampled at the measurement interval, which the stride of one period collapses.
The two computations contain the same information, but the period stride needs a rank equal to the number of excited multipliers rather than of excited harmonics (two for the Mathieu equation, where Hankel-DMD needed sixteen for the same accuracy), represents the entire periodic base motion by the single eigenvalue one, and returns directly the quantity by which Floquet theory decides stability; standard DMD, the window of one sample, finds the comb as sidebands only when the harmonics are spatially independent.
On measurements from a combustor, DMD and FD, which see one snapshot at a time, miss the two modes associated with the instability, while EFD and Hankel-DMD find both, with matching eigenvalues and modes, EFD more consistently and Hankel-DMD only with a window of at least about one analysis period, so that knowing the period gives Hankel-DMD its shortest useful window; these are the modes of the acoustically coupled combustion instability that the companion paper suggests is parametric \cite{Quinlan-2026-PCI}.

The verification also marks the limits of the method.
Near a limit cycle the transverse multiplier is recovered in proportion to the perturbation, and the powers of the multiplier that the nonlinearity generates can be absorbed by raising the rank, for which EFD leaves room when the observables are few.
For data far from a linear periodic system, a conservative nonlinear oscillator and the measurements from a turbulent combustor, the arguments of the eigenvalues remain informative but the moduli are biased toward the origin, and a modulus below one does not establish decay.
The period must be estimated, and frequencies are insensitive to an error of one sample in ten when a delay window and spare rank are used; the method does not resolve the spectrum within a period beyond the phase dependence of its modes, and it does not apply to aperiodic time dependence, for which methods based on local estimation \cite{Macesic-2018,Zhang-2019} are designed.
A noise analysis of FD and EFD, including the bias of the moduli and the benefit of the stacked regression, is the most important open problem for their use on measured data.

FD and EFD require a common period, which a system modulated at two incommensurate frequencies does not possess.
Natural extensions are FD over sliding windows when the second frequency is slow; a continued-fraction convergent of the frequency ratio as the analysis period when the ratio is close to rational, as it is exactly when nonlinear oscillators lock; and, for reducible quasi-periodic systems \cite{Johnson-1981,Jorba-1992}, delay-embedded DMD of one trajectory, which should return each quasi-Floquet exponent replicated on the lattice of Mezi\'c and Surana \cite[Theorem~5.1]{Mezic-2016}, the analog of the comb of \cref{sec:hankel}, so that grouping the eigenvalues into lattice families would recover the exponents without any period.
Where the phases of the modulating processes can be measured, a regression conditioned on phase over the torus would generalize EFD to two drivers.
In all of these the Floquet multiplier gives way to Lyapunov exponents or the Sacker--Sell spectrum \cite{Sacker-1978} as the measure of stability.

%% file: sections/A_proofs.tex
\section{Formal statements and proofs}
\label{app:proofs}

This appendix states the results of \cref{sec:methods,sec:relations} formally and proves them.
Each result is preceded by a sentence that says what it means, and each proof by the idea on which it rests, so that the argument can be followed without the details; the details are there for the reader who wishes to check them.
The notation is that of \cref{sec:floquet}: the data satisfy the model \cref{eq:sampled} with commensurate sampling \cref{eq:commensurate}, $\mM_i$ is the monodromy matrix at phase $i$ \cref{eq:Mi}, $\mObs_k^{(d)}$ is the observability matrix over a window \cref{eq:delay-vector}, and $\Kry_i$ and $\nt$ are defined in \cref{eq:Ki}.

\subsection{DMD of an observed Krylov sequence}
\label{app:krylov}

Everything else rests on one fact about DMD, \cref{lem:krylov}.
It concerns vectors produced by applying one matrix $\mF$ repeatedly to a starting vector $\vz_1$ and measuring each result through a matrix $\mO$.
The iterates $\vz_1,\mF\vz_1,\mF^2\vz_1,\dots$ span the \emph{Krylov subspace} $\Kry$ of $\mF$ and $\vz_1$, the smallest subspace that contains $\vz_1$ and that $\mF$ maps into itself; its dimension $r$ is the number of eigenvalues of $\mF$ that $\vz_1$ excites.
The measurement must be \emph{one-to-one} (injective) on $\Kry$, which means that no non-zero vector of $\Kry$ is mapped to zero, so that different states in $\Kry$ give different data.
The lemma then says that DMD of rank $r$ returns exactly the eigenvalues of $\mF$ that $\vz_1$ excites, and their eigenvectors as they are seen through $\mO$.
It is phrased with two further terms.
The \emph{restriction} $\mF|_{\Kry}$ is $\mF$ regarded as a map of $\Kry$ into itself, whose eigenvalues are those of $\mF$ with eigenvectors in $\Kry$.
Two square matrices $\mS$ and $\widehat{\mF}$ are \emph{similar} if $\mS = \mTheta^{-1}\widehat{\mF}\mTheta$ for some invertible matrix $\mTheta$; they describe the same map in two systems of coordinates, and they have the same eigenvalues.

\begin{lemma}[DMD of an observed Krylov sequence]
\label{lem:krylov}
Let $\mF\in\CC^{q\times q}$ and $\vz_1\in\CC^q$, let $\vz_{k+1} = \mF\vz_k$ for $k\ge 1$, let $\Kry = \spn\{\vz_1,\vz_2,\dots\}$ and $r = \dim\Kry$, and let $\mO\in\CC^{\tilde{m}\times q}$ be one-to-one on $\Kry$.
For $K-1\ge r$ set
\begin{equation*}
  \Vx = \mO\begin{bmatrix} \vz_1 & \vz_2 & \cdots & \vz_{K-1} \end{bmatrix},
  \qquad
  \Vy = \mO\begin{bmatrix} \vz_2 & \vz_3 & \cdots & \vz_K \end{bmatrix}.
\end{equation*}
Then $\rank\Vx = r$, and the rank-$r$ DMD matrix $\mS$ of \cref{eq:Stilde} is similar to the restriction $\mF|_{\Kry}$.
In particular the DMD eigenvalues are the eigenvalues of $\mF|_{\Kry}$, with their multiplicities, and if $\mF\vw=\lambda\vw$ with $\vw\in\Kry$ then $\mO\vw$ is the DMD mode for $\lambda$.
\end{lemma}

\emph{Idea of the proof.}
Each data matrix factors into a measurement part, the images under $\mO$ of a basis of $\Kry$, and an iteration part, the coordinates of the iterates in that basis.
The pseudo-inverse undoes both factors, and what remains of the fitted matrix is $\mF$ itself, acting on the measured directions as it acts on $\Kry$.

\begin{proof}
Let the columns of $\mQ\in\CC^{q\times r}$ be a basis of $\Kry$.
Since $\mF$ maps $\Kry$ into itself, there is a unique $\widehat{\mF}\in\CC^{r\times r}$, the matrix of $\mF|_{\Kry}$ in this basis, with $\mF\mQ = \mQ\widehat{\mF}$.
Write $\vz_1 = \mQ\vc$, so that $\vz_k = \mF^{k-1}\vz_1 = \mQ\widehat{\mF}^{k-1}\vc$ and $[\,\vz_1\ \cdots\ \vz_{K-1}\,] = \mQ\mR$ with $\mR = [\,\vc\ \ \widehat{\mF}\vc\ \cdots\ \widehat{\mF}^{K-2}\vc\,]\in\CC^{r\times (K-1)}$.
The first $r$ vectors of a Krylov sequence span its Krylov subspace, so $\mR$ has full row rank (linearly independent rows) for $K-1\ge r$.
Set $\mG = \mO\mQ$, which has full column rank (linearly independent columns) because $\mO$ is one-to-one on $\Kry$.
Then $\Vx = \mG\mR$ and $\Vy = \mG\widehat{\mF}\mR$, so $\rank\Vx = r$, $\Vx^{\pinv} = \mR^{\pinv}\mG^{\pinv}$, and
\begin{equation}
  \mA = \Vy\Vx^{\pinv} = \mG\widehat{\mF}\,\mR\mR^{\pinv}\mG^{\pinv} = \mG\,\widehat{\mF}\,\mG^{\pinv},
  \label{eq:A-GFG}
\end{equation}
because $\mR\mR^{\pinv} = \mI$ for a matrix with full row rank.
The matrices $\mU$, $\Vx$, and $\mG$ have the same range (column space), so the matrix $\mTheta = \mG^{\pinv}\mU$ satisfies $\mG\mTheta = \mU$ and is invertible, with $\mU^{\herm}\mG = \mTheta^{-1}$.
Hence $\mS = \mU^{\herm}\mA\mU = \mU^{\herm}\mG\,\widehat{\mF}\,\mG^{\pinv}\mU = \mTheta^{-1}\widehat{\mF}\,\mTheta$.
Finally, if $\mF\vw = \lambda\vw$ with $\vw = \mQ\vb{b}$, then $\widehat{\mF}\vb{b} = \lambda\vb{b}$ and $\mA(\mO\vw) = \mG\widehat{\mF}\mG^{\pinv}\mG\vb{b} = \lambda\,\mO\vw$.
\end{proof}

\subsection{FD and EFD}
\label{app:fd-efd}

FD and EFD are \cref{lem:krylov} with the monodromy matrix in the part of $\mF$.
They differ in the measurement: FD sees the state through $\mC_i$ at one phase, and EFD through the stacked matrix $\mObs_1^{(p)}$, which measures the state at every phase of one period.
The requirement that the measurement be one-to-one is therefore a requirement on one instant for FD and on one period for EFD.

\begin{proposition}[exactness of FD]
\label{prop:fd}
Let the data satisfy \cref{eq:sampled}.
If $\mC_i$ is one-to-one on $\Kry_i$ and $K>\nt$, then the FD of rank $\nt$ returns exactly the eigenvalues of $\mM_i|_{\Kry_i}$.
These are Floquet multipliers of \cref{eq:ltp}, and they are all of them when $\nt=n$.
The mode returned with $\mu_l$ is $\mC_i \vw_l^{(i)} = \ee^{\rho_l t_i}\,\mC_i\,\vphi_l(t_i)$, the observed Floquet mode at phase $t_i$.
\end{proposition}

\begin{proof}
By \cref{eq:period-map}, column $k$ of \cref{eq:FD-data} is $\snap_{i+(k-1)p} = \mC_i\mM_i^{\,k-1}\vx_i$.
Apply \cref{lem:krylov} with $\mF = \mM_i$, $\vz_1 = \vx_i$, and $\mO = \mC_i$, and use \cref{eq:Mi-eigvec} for the eigenvectors.
\end{proof}

\begin{proposition}[exactness of EFD]
\label{prop:efd}
Let the data satisfy \cref{eq:sampled}.
If the system is observable over one period on $\Kry_1$, that is, if $\mObs_1^{(p)}$ is one-to-one on $\Kry_1$, and $K>\nt$, then
\begin{enumerate}
\item[(a)] $\rank\Vxt = \nt$ for the stacked matrix $\Vxt$ of \cref{eq:EFD-data};
\item[(b)] the EFD of rank $\nt$ returns exactly the eigenvalues of $\mM_1|_{\Kry_1}$, which are Floquet multipliers of \cref{eq:ltp};
\item[(c)] the mode returned with $\mu_l$ is $\vpsit_l = \mObs_1^{(p)}\vw_l^{(1)}$, whose $c$-th block is given by \cref{eq:efd-mode-block}.
\end{enumerate}
\end{proposition}

\begin{proof}
By \cref{eq:h-equals-Ox,eq:period-map}, $\stk_{(k-1)p+1}^{(p)} = \mObs_1^{(p)}\vx_{(k-1)p+1} = \mObs_1^{(p)}\mM_1^{\,k-1}\vx_1$.
Apply \cref{lem:krylov} with $\mF = \mM_1$, $\vz_1 = \vx_1$, and $\mO = \mObs_1^{(p)}$; the blocks of $\mObs_1^{(p)}\vw_l^{(1)}$ are given by \cref{eq:delay-vector,eq:Mi-eigvec,eq:floquet-mode}.
\end{proof}

The next statement says that removing the known factor $\ee^{\rho_l t_c}$ from each block of an EFD mode, and taking a discrete Fourier transform across the blocks, gives the harmonics of the observed Floquet mode.

\begin{proposition}[phase alignment and harmonics]
\label{prop:harmonics}
Under the hypotheses of \cref{prop:efd}, the identity \cref{eq:harmonic} holds for every integer $j$: the transform across the aligned blocks of the EFD mode returns the discrete Fourier coefficients $\vpsih_{l,j}$ of the observed Floquet mode.
\end{proposition}

\begin{proof}
Substitute \cref{eq:efd-mode-block} and use $\omega t_c = 2\pi c/p$.
\end{proof}

The statement of \cref{sec:base-state} about a periodic base state follows in the same way.
The stacked columns are $\bar{\stk} + \mObs_1^{(p)}\mM_1^{\,k-1}\xp_1$, $k = 1,\dots,K$.
A constant vector is an iterate of the number one, so they are an observed Krylov sequence of the matrix $\blkdiag(1,\mM_1)$ started from $(1,\xp_1)$, and \cref{lem:krylov} applies when $[\,\bar{\stk}\ \ \mObs_1^{(p)}\,]$ is one-to-one on the corresponding Krylov subspace.

\subsection{The cyclic reformulation and Hankel-DMD}
\label{app:hankel}

Hankel-DMD advances the data by one sample, and from one sample to the next the system is not time-invariant.
The \emph{cyclic reformulation} makes it time-invariant by enlarging the state.
The enlarged state has $p$ blocks, one for each phase of the period.
At sample $k$ the state $\vx_k$ sits in the block of its phase, the block $i\in\{1,\dots,p\}$ for which $k-i$ is a multiple of $p$, and the other blocks are zero; one step of a constant matrix $\mF$ moves it to the next block, from block $p$ back to block $1$, while advancing it by $\mPhi_k$.
After $p$ steps the state is back in its own block, advanced by one period, which is part (b) of \cref{lem:cyclic}.
Part (c) says that the eigenvalues of $\mF$ are the $p$-th roots of the multipliers, the teeth of the comb of \cref{sec:sidebands}.

\begin{lemma}[cyclic reformulation]
\label{lem:cyclic}
For the system \cref{eq:sampled} define
\begin{equation}
  \mF = \begin{bmatrix}
        \mat{0} & \mat{0} & \cdots & \mat{0} & \mPhi_p \\
        \mPhi_1 & \mat{0} & \cdots & \mat{0} & \mat{0} \\
        \mat{0} & \mPhi_2 & \ddots & \vdots & \vdots \\
        \vdots & & \ddots & \mat{0} & \mat{0} \\
        \mat{0} & \mat{0} & \cdots & \mPhi_{p-1} & \mat{0}
      \end{bmatrix} \in \CC^{np\times np},
  \label{eq:cyclic-F}
\end{equation}
let $\boldsymbol{\mathcal{C}} = \bigl[\, \mC_1 \;\; \mC_2 \;\; \cdots \;\; \mC_p \,\bigr] \in \CC^{m\times np}$, and let $\vz_k\in\CC^{np}$ have $\vx_k$ in the block $i\in\{1,\dots,p\}$ for which $k-i$ is a multiple of $p$, and zeros elsewhere.
Then
\begin{enumerate}
\item[(a)] $\vz_{k+1} = \mF\vz_k$ and $\snap_k = \boldsymbol{\mathcal{C}}\vz_k$ for all $k\ge 0$;
\item[(b)] $\mF^p = \blkdiag(\mM_1,\mM_2,\dots,\mM_p)$;
\item[(c)] $\beta$ is an eigenvalue of $\mF$ if and only if $\beta^p$ is a Floquet multiplier.
If $\mM\vw = \mu\vw$ and $\beta^p = \mu$, an eigenvector of $\mF$ for $\beta$ is the vector with blocks $\beta^{-i}\,\mPhi(t_i,0)\,\vw$, $i=1,\dots,p$.
\end{enumerate}
\end{lemma}

\begin{proof}
$\mF$ maps a vector with $\vx$ in block $i$ to the vector with $\mPhi_i\vx$ in the next block, block $1$ following block $p$, and $\boldsymbol{\mathcal{C}}$ maps it to $\mC_i\vx$; since $\mPhi_{k+p} = \mPhi_k$ and $\mC_{k+p} = \mC_k$, this gives (a), and $p$ applications give (b) by \cref{eq:Mi}.
For (c), let $\mF\vz = \beta\vz$ with blocks $\vz_{[i]}$.
Then $\mPhi_i\vz_{[i]} = \beta\vz_{[i+1]}$ for $i<p$ and $\mPhi_p\vz_{[p]} = \beta\vz_{[1]}$, and $\beta\ne 0$ because $\mF$ is invertible, as every $\mPhi_i$ is.
Hence $\vz_{[p]}\ne \vb{0}$, since otherwise all blocks would vanish, and $\mM_p\vz_{[p]} = \mPhi_{p-1}\cdots\mPhi_1\mPhi_p\vz_{[p]} = \beta^p\vz_{[p]}$: the number $\beta^p$ is an eigenvalue of $\mM_p = \mM$.
Conversely, for the stated vector, $\mPhi_i\,\beta^{-i}\mPhi(t_i,0)\vw = \beta\cdot\beta^{-(i+1)}\mPhi(t_{i+1},0)\vw$ for $i<p$, and $\mPhi_p\,\beta^{-p}\mPhi(t_p,0)\vw = \beta^{-p}\mPhi(t_1,0)\mM\vw = \beta\cdot\beta^{-1}\mPhi(t_1,0)\vw$, because $\mPhi(t_{p+1},0) = \mPhi(t_1+T,T)\mPhi(T,0) = \mPhi(t_1,0)\mM$ and $\beta^{-p}\mu = 1$.
\end{proof}

Parts (a) and (b) are the classical reformulation \cite{Meyer-1975,Bittanti-2009}.
For the remainder of this appendix assume that $\mM$ is diagonalizable, with eigenpairs $(\mu_l,\vw_l)$ and $\vx_0 = \sum_la_l\vw_l$, and let $\beta_{l,j}$, $\zeta$, $\vpsih_{l,j}$, $\Jset$, $r$, and $\vb{v}_d(\beta)$ be as in \cref{eq:teeth,eq:harmonic,eq:Jset,eq:h-comb}.
The next lemma says that the snapshots are a sum of pure exponentials $\beta_{l,j}^{\,k}$ with fixed shapes $\vpsih_{l,j}$, the carrier and sidebands of \cref{sec:sidebands}; it is proved here through the eigenvectors of the cyclic matrix.

\begin{lemma}[comb expansion]
\label{lem:comb}
The expansion \cref{eq:comb-expansion} holds for all $k\ge 0$.
\end{lemma}

\begin{proof}
By \cref{lem:cyclic}(c) the vector $\vz^{(l,j)}$ with blocks $\beta_{l,j}^{-i}\mPhi(t_i,0)\vw_l$ is an eigenvector of $\mF$ for $\beta_{l,j}$.
Block $i$ of $\sum_j\vz^{(l,j)}$ equals $\beta_l^{-i}\mPhi(t_i,0)\vw_l\sum_j\zeta^{-ij}$, which is $\mu_l^{-1}\mM\vw_l\,p = p\,\vw_l$ for $i=p$ and zero otherwise, because the $p$-th roots of unity sum to zero.
The initial state is at the initial phase, which is phase $p$, so $\vz_0$ has $\vx_0$ in block $p$; therefore $\vz_0 = p^{-1}\sum_la_l\sum_j\vz^{(l,j)}$, and by \cref{lem:cyclic}(a),
\begin{equation*}
  \snap_k = \boldsymbol{\mathcal{C}}\mF^k\vz_0 = \frac{1}{p}\sum_{l,j}a_l\,\beta_{l,j}^{\,k}\,\boldsymbol{\mathcal{C}}\vz^{(l,j)} .
\end{equation*}
Finally, by \cref{eq:floquet-mode,eq:harmonic},
\begin{equation*}
  \boldsymbol{\mathcal{C}}\vz^{(l,j)} = \sum_i\zeta^{-ij}\,\beta_l^{-i}\,\mC_i\mPhi(t_i,0)\vw_l = \sum_i\zeta^{-ij}\,\vpsi_l(t_i) = p\,\vpsih_{l,j} .
\end{equation*}
\end{proof}

Every eigenvector of the cyclic matrix is excited by a trajectory that starts at the initial time, each tooth $\beta_{l,j}$ with the same weight $a_l/p$; whether a tooth is seen in the output is decided by $\vpsih_{l,j}$.
The derivation of \cref{sec:sidebands}, which substitutes the discrete Fourier series of the observed Floquet mode into the Floquet expansion, is an equivalent proof.

The theorem on Hankel-DMD is \cref{lem:krylov} once more.
By \cref{lem:comb} the stacked snapshots are iterates of the diagonal matrix of the teeth, measured through the matrix whose columns are the stacked harmonics; a long enough window makes that measurement one-to-one, and DMD then returns the teeth.
The proof uses a \emph{Vandermonde matrix}, whose rows are the successive powers $(1,\beta,\beta^2,\dots)$ of a set of numbers $\beta$, one row for each number; when the numbers are distinct and there are at least as many columns as rows, its rows are linearly independent, and so are the columns of its transpose.

\begin{theorem}[Hankel-DMD of a linear periodic system]
\label{thm:hankel}
Let the data satisfy \cref{eq:sampled} with commensurate sampling \cref{eq:commensurate}, and let the Floquet multipliers be distinct.
Let $(\Vx,\Vy)$ be the stride-one pair \cref{eq:XY} with window $d$ and $K$ columns.
Then $\rank\Vx\le r$, with equality whenever $d\ge r$ and $K>r$.
If $\rank\Vx = r$, the Hankel-DMD of rank $r$ has the following properties.
\begin{enumerate}
\item[(a)] Its eigenvalues are exactly the teeth $\beta_{l,j}$, $(l,j)\in\Jset$.
Each is a $p$-th root of a Floquet multiplier, $\beta_{l,j}^{\,p} = \mu_l$.
\item[(b)] For each multiplier excited by the initial state, the number of teeth returned equals the number of non-zero discrete harmonics of the observed Floquet mode $\vpsi_l$.
It is at most $p$, and at least one unless the mode is invisible to the measurement at every phase.
\item[(c)] The mode returned with $\beta_{l,j}$ is $\vb{v}_d(\beta_{l,j})\otimes\vpsih_{l,j}$; its leading block is the $j$-th harmonic $\vpsih_{l,j}$ of the observed Floquet mode.
\end{enumerate}
\end{theorem}

\begin{proof}
Index the elements of $\Jset$ by $\sigma = 1,\dots,r$ and write $\beta_\sigma$, $\vpsih_\sigma$, $a_\sigma$ accordingly.
By \cref{eq:comb-expansion}, $\stk_k^{(d)} = \sum_\sigma\beta_\sigma^{\,k}\vg_\sigma$ with $\vg_\sigma = a_\sigma\,\vb{v}_d(\beta_\sigma)\otimes\vpsih_\sigma$.
With $\mG = [\,\vg_1\ \cdots\ \vg_r\,]\in\CC^{md\times r}$, $\mD = \diag(\beta_\sigma)$, and $\vone$ the vector of ones, this reads $\stk_k^{(d)} = \mG\,\mD^k\vone$, an observed Krylov sequence of $\mD$ started from $\mD\vone$ at $k=1$.
Hence $\Vx = \mG\mD\,[\,\vone\ \ \mD\vone\ \cdots\ \mD^{K-2}\vone\,]$ and $\rank\Vx\le r$.
The $\beta_\sigma$ are distinct: $\beta_{l,j} = \beta_{l',j'}$ implies $\mu_l = \mu_{l'}$, hence $l=l'$, and then $j=j'$ modulo $p$.
The matrix $[\,\vone\ \ \mD\vone\ \cdots\ \mD^{K-2}\vone\,]$, whose row $\sigma$ is $(1,\beta_\sigma,\dots,\beta_\sigma^{K-2})$, is therefore a Vandermonde matrix with distinct numbers, of rank $r$ when $K-1\ge r$; since $\mD$ is invertible, the Krylov subspace of $\mD$ and $\mD\vone$ is then all of $\CC^r$.
If $d\ge r$ the vectors $\vb{v}_d(\beta_\sigma)$ are linearly independent for the same reason, and then so are the $\vg_\sigma$.
To see this, suppose that $\sum_\sigma\alpha_\sigma a_\sigma \vb{v}_d(\beta_\sigma)\otimes\vpsih_\sigma = \vb{0}$.
Collecting the $q$-th entry of every block gives $\sum_\sigma\alpha_\sigma a_\sigma(\vpsih_\sigma)_q\,\vb{v}_d(\beta_\sigma) = \vb{0}$ for each $q$, so that $\alpha_\sigma a_\sigma(\vpsih_\sigma)_q = 0$ for all $\sigma$ and $q$; since $a_\sigma\vpsih_\sigma\ne \vb{0}$, every $\alpha_\sigma$ vanishes.
This proves the statements about the rank.
Conversely, $\rank\Vx = r$ forces $\mG$ to have full column rank and $K>r$.
\Cref{lem:krylov} with $\mF = \mD$, $\vz_1 = \mD\vone$, and $\mO = \mG$ then shows that $\mS$ is similar to $\mD$ and that the mode for $\beta_\sigma$ is $\vg_\sigma$, which is (a) and (c).
Part (b) restates the definition of $\Jset$.
\end{proof}

The condition $\rank\Vx = r$ is, in terms of \cref{lem:cyclic}, the requirement that the part of the cyclic system that the trajectory excites be observable over the window.
The assumption of distinct multipliers serves the explicit description by harmonics.
Without it the explicit description is lost, but the conclusion about the eigenvalues survives.
Restricted to the states that the trajectory excites and that the measurement over the window can tell apart, the cyclic system is a constant system observed through a one-to-one measurement, and an argument like that of \cref{lem:krylov} shows that every Hankel-DMD eigenvalue is an eigenvalue of $\mF$ once the rank of $\Vx$ has stopped growing with $d$ and $K$, and hence a $p$-th root of a multiplier by \cref{lem:cyclic}(c).

The last two results compare strides.
Taking every $p$-th column raises each tooth to the power $p$, which sends all the teeth of one multiplier to the multiplier itself; they then advance together and merge into a single eigenvalue, whose mode is the sum of their modes.
\Cref{cor:collapse} states this for the fitted matrices.
It uses the \emph{orthogonal projector} onto a subspace, the matrix that leaves the vectors of the subspace unchanged and sends the vectors perpendicular to it to zero.

\begin{corollary}[EFD and Hankel-DMD]
\label{cor:collapse}
Assume the hypotheses of \cref{thm:hankel} with $\rank\Vx = r$, let $\mA_1 = \Vy\Vx^{\pinv}$ be the stride-one operator, and let $\mA_p = \Vyt\Vxt^{\pinv}$ be the operator obtained from the same delay vectors with stride $p$ and at least $n_{\Jset}$ column pairs, where $n_{\Jset}$ is the number of multipliers that have a tooth in $\Jset$.
Then
\begin{enumerate}
\item[(a)] $\range\Vxt\subseteq\range\Vx$ is invariant under $\mA_1^{\,p}$, that is, mapped into itself, and $\mA_p = \mA_1^{\,p}\,\mPi_p$, where $\mPi_p$ is the orthogonal projector onto the range of $\Vxt$;
\item[(b)] the non-zero eigenvalues of $\mA_1^{\,p}$ are these multipliers $\mu_l$, each repeated once for every tooth $(l,j)\in\Jset$, while the non-zero eigenvalues of $\mA_p$ are the same multipliers, each once;
\item[(c)] the mode of $\mA_p$ for $\mu_l$ is the sum $\vpsit_l$ of \cref{eq:mode-sum} of the Hankel-DMD modes over the comb of $\mu_l$, each multiplied by its tooth, and conversely the $j$-th harmonic \cref{eq:harmonic} of the phase-aligned EFD mode is the leading block of the Hankel-DMD mode at $\beta_{l,j}$.
\end{enumerate}
\end{corollary}

\begin{proof}
We use the notation of the previous proof, in which $\mD^p = \diag(\mu_{l(\sigma)})$ and the $k$-th column at stride $p$ is $\stk_{(k-1)p+1}^{(d)} = \mG\,\mD^{(k-1)p}\mD\vone$.
Let $\mE\in\{0,1\}^{r\times n_{\Jset}}$ have a one in position $(\sigma,l)$ when tooth $\sigma$ belongs to multiplier $l$, and let $\mLambda = \diag(\mu_l)$ over the multipliers that have a tooth in $\Jset$.
Then $\mD^p\mE = \mE\mLambda$ and $\vone = \mE\vone$, so $\stk_{(k-1)p+1}^{(d)} = \mG_p\mLambda^{k-1}\vone$ with $\mG_p = \mG\mD\mE$, which has full column rank.
\Cref{lem:krylov}, as in \cref{eq:A-GFG}, gives $\mA_p = \mG_p\mLambda\mG_p^{\pinv}$ and $\mA_1 = \mG\mD\mG^{\pinv}$, so $\mA_1^{\,p} = \mG\mD^p\mG^{\pinv}$ because $\mG^{\pinv}\mG = \mI$.
Then $\mA_1^{\,p}\mG_p = \mG\mD^{p+1}\mE = \mG\mD\mE\mLambda = \mG_p\mLambda = \mA_p\mG_p$: the two operators agree on $\range\mG_p = \range\Vxt$, which is invariant under both.
$\mA_p$ vanishes on the orthogonal complement of $\range\mG_p$ (the vectors perpendicular to it), which proves (a), and (b) follows from the two factorizations.
The columns of $\mG_p$ are $a_l\vpsit_l$ with $\vpsit_l$ as in \cref{eq:mode-sum}, and the inverse discrete Fourier transform shows that block $c$ of $\vpsit_l$ is $\ee^{\rho_lt_c}\vpsi_l(t_c)$, in agreement with \cref{eq:efd-mode-block}.
The last statement is \cref{thm:hankel}(c) combined with \cref{prop:harmonics}.
\end{proof}

Standard DMD is the case of a window of one sample.
The window can then no longer separate the teeth, and they are returned only if their spatial shapes are linearly independent.

\begin{corollary}[standard DMD]
\label{cor:standard}
Under the hypotheses of \cref{thm:hankel}, standard DMD returns the teeth $\beta_{l,j}$, $(l,j)\in\Jset$, if and only if the $r$ vectors $a_l\vpsih_{l,j}\in\CC^m$ are linearly independent and $K>r$.
This requires $m\ge r$.
\end{corollary}

\begin{proof}
For $d=1$ the matrix $\mG$ in the proof of \cref{thm:hankel} has the columns $a_l\vpsih_{l,j}$.
\end{proof}